\documentclass[11pt,a4paper]{article}
\usepackage[T1]{fontenc}
\usepackage{lmodern}
\usepackage{tikz}
\usepackage{graphicx} 
\usepackage{amstext} 
\usepackage{amsmath} 
\usepackage[top=25mm,bottom=37mm,left=20mm,right=20mm]{geometry} 
\usepackage{color} 
\definecolor{myblu}{rgb}{0.1,0.1,0.5}
\usepackage{hyperref} 
\hypersetup{colorlinks=true,urlcolor=blue,linkcolor=black,citecolor=black} 
\usepackage{sectsty,textcase} 
\sectionfont{\large\MakeTextUppercase} 
\usepackage{secdot} 
\usepackage{amssymb}
\usepackage[backend=bibtex,style=numeric,sorting=none,maxbibnames=5]{biblatex}

\bibliography{biblio}

\everymath{\displaystyle}
\begin{document}
\title{\vspace{-18mm}
\textbf{\Large A priori tests of turbulence models for compressible flows}}
 \author{\textbf{L. Sciacovelli$^\text{(1)}$, A. Cannici$^\text{(2)}$, D. Passiatore$^\text{(3)}$ and P. Cinnella$^\text{(4)}$}\\
 {\normalsize\itshape
 $^\text{(1)}$Arts et Métiers, DynFluid Laboratory, Paris France, luca.sciacovelli@ensam.eu}
 \\{\normalsize\itshape
 $^\text{(2)}$Arts et Métiers, DynFluid Laboratory, Paris France, aron.cannici@gmail.com}
 \\{\normalsize\itshape
 $^\text{(3)}$Stanford University, Center for Turbulence Research,  Stanford, CA, USA, dodipass@stanford.edu}
 \\{\normalsize\itshape
 $^\text{(4)}$Sorbonne Université, Institut Jean le Rond d’Alembert, Paris France, paola.cinnella@sorbonne-universite.fr}}
\maketitle

\begin{abstract}
A priori tests of turbulence models for the compressible Reynolds-Averaged Navier--Stokes (RANS) are performed by using Direct Numerical Simulations (DNS) data of zero-pressure-gradient flat-plate turbulent boundary layers.
The DNS database covers a wide range of operating conditions, ranging from supersonic ($M_\infty = 2.25$) up to the high-enthalpy hypersonic regime ($M_\infty= 12.48$).
Several RANS closures and compressibility corrections in the literature are assessed against the exact terms from the DNS. Particular attention is paid to closure models for the turbulent heat fluxes and the dilatational dissipation, as well as to the analysis of turbulence/chemistry and turbulence/vibrational relaxation interactions for the high-enthalpy simulations.
\end{abstract}

\section{Introduction}
Accurate modeling of high-speed flows is a cornerstone in the design of novel aerospace components that ensure the safe re-entry of spacecraft, advance reusable launch vehicles, and pioneer suborbital flights for rapid civilian transportation, collectively reshaping the realms of space exploration, travel, and technology. More specifically, vehicles traveling at supersonic and hypersonic speeds experience extreme mechanical and thermal loads due to aerodynamic effects.
As the flow transitions to turbulence, a significant increase in skin friction and wall heat fluxes occurs, which must be accurately accounted into the design of propellers and thermal protection systems.
In general, numerical simulations of turbulent flows for industrial applications rely on Reynolds-Averaged Navier-Stokes (RANS) models, most of which are designed and calibrated for incompressible flows.
However, when dealing with compressible flows, additional unclosed terms related to the variable density and transport properties appear due to the Favre averaging of the Navier-Stokes equations. Such terms are not present in the RANS equations for incompressible, constant--property flows, and are very often neglected in compressible flow models.
While this choice provides acceptable results at low supersonic Mach numbers, it becomes unsuitable when compressibility and/or non-equilibrium effects become relevant. Attempts to develop RANS closures well-suited to the compressible regime exist in the literature, but they are restricted to relatively-simple canonical configurations, such as  homogeneous isotropic turbulence or mixing layers. Relevant examples in the literature are the works of Sarkar \emph{et al.}~\cite{sarkar1991analysis} and Zeman~\cite{zeman1990dilatation}, who introduced the concept of dilatational dissipation for compressible turbulence and incorporated it into a closure model for mixing layers. Later, Wilcox~\cite{wilcox1992dilatation} extended this concept to flat-plate boundary layers and proposed a modification of the $k$-$\omega$ model for compressible flows. Catris and Aupoix~\cite{catris2000density} introduced a density scaling in the diffusion terms of the turbulent transport equations to account for variable flow properties across high-speed boundary layers due to frictional heating. More recently, Pecnik and Patel~\cite{Pecnik2017scaling} investigated variable-property scaling for non-adiabatic, low-Mach turbulent duct flows and obtained a new scaling of the diffusion terms \cite{otero2018turbulence} that is quite similar to that of Catris and Aupoix~\cite{catris2000density}.
A comprehensive evaluation of turbulence models for compressible turbulent flows was performed in the review paper by Roy and Blottner \cite{roy2006review}, which shows that while some of the turbulence models provide reasonable predictions for surface pressure, the predictions for surface heat flux are generally poor, often off by a factor of four or more.
Overall, closure models for the compressible RANS equations remain an open subject of research. The closure problem is even more complex for high-enthalpy regimes, where thermochemical non-equilibrium effects such as dissociation/recombination and vibrational relaxation processes may play an important role in the overall flow dynamics and have to be accounted for. The modeling of the additional unclosed terms for turbulent transport of chemical species or vibrational energy introduces further uncertainties in RANS approaches, which are not well-quantified yet, due to the lack of reference data.\\
Performing physical experiments in these extreme flow conditions is generally a costly or even infeasible task, whence the interest of leveraging high-fidelity approaches and specifically Direct Numerical Simulations (DNS).
DNS is computationally expensive and usually limited to simple geometries and low Reynolds numbers, and this is even more true for high-enthalpy hypersonic flows, for which additional transport equations for the chemical species in the reacting mixture and for thermal non-equilibrium must be solved alongside the standard equations for mass, momentum, and total energy. Despite these limitations, DNS represents an invaluable resource for fundamental understanding of turbulence dynamics and for the assessment of RANS models.
Recently, our team has generated some of the very first DNS for high-enthalpy hypersonic flows out of chemical or thermochemical equilibrium \cite{passiatore2021finite,passiatore2022thermochemical,passiatore2023shock}. The results showed the presence of a marked interaction between turbulence and non-equilibrium conditions; specifically, vibrational relaxation and chemical dissociation processes are promoted and sustained by the onset of turbulent motions. As a consequence, the classical approach of neglecting turbulence/thermochemistry interactions underestimates the source terms for the chemical production rates and the translational-vibrational energy exchanges. In addition, the lack of properly designed closures for the turbulent fluxes of species and vibrational energy equations lead us to investigate the behavior of the existing simplified models for these configurations, motivating the present study.\\

In this work, we assess the validity of closure models for the compressible RANS equations by performing \emph{a priori} tests. For that purpose, we use a database of several DNS of zero-pressure-gradient flat-plate turbulent boundary layers (TBL) ranging from supersonic to high-enthalpy hypersonic conditions \cite{sciacovelli2021assessment,passiatore2021finite,passiatore2022thermochemical} to compare the exact and modeled counterparts of several unclosed terms appearing in the averaged momentum, energy, and thermo-chemistry transport equations. Several models from the literature employed in compressible RANS codes are scrutinized.\\
The structure of the paper is as follows: the numerical databases are described in section~\ref{sec:method}; section~\ref{sec:models} presents the averaged equations and the main closure models investigated. Results are then discussed in section~\ref{sec:results} and conclusive remarks and perspectives are given in section~\ref{sec:conclusions}.
\section{Numerical Databases}\label{sec:method}

\begin{table*}[!tb]
\centering
\caption{Parameters of the DNS considered for the tests. $M_\infty$, $T_\infty$ and $P_\infty$ denote the free-stream Mach number, temperature and pressure, respectively. $T_w$ lists the isothermal/adiabatic wall boundary condition for each case; $N_x$, $N_y$ and $N_z$ represent the number of grid points in the streamwise, wall-normal and spanwise direction; $L_x$, $L_y$ and $L_z$ are the extents of the computational domains, $\delta_\text{in}^*$ being the displacement thickness at the inflow section.}
\vspace{2mm}
\begin{tabular}{| c c c c c c c c c c c |}
\hline
Case & $M_\infty$ & $T_\infty$ [K] & $P_\infty$ [Pa] & $T_w$ [K] & $N_x$ & $N_y$ & $N_z$ & $L_x$ & $L_y$ & $L_z$ \\ [0.5ex]
\hline\hline
M2 \cite{sciacovelli2021assessment}      & 2.25  & 65   & 2422 & $120.2$ &  8000 & 300 & 512 & 1600 $\delta_\text{in}^{\star}$ & 100 $\delta_\text{in}^{\star}$ & 62.83 $\pi \delta_\text{in}^{\star}$  \\
M6  \cite{sciacovelli2020numerical}   & 6     & 78   & 2422 & $422.5$ &  7700 & 300 & 400 & 75.2 $\delta_{99}^\text{out}$ & 2.13 $\delta_{99}^\text{out}$ & 2.17 $\delta_{99}^\text{out}$  \\
M10F    & 10    & 350  & 3596 &  Adiab. &  5520 & 256 & 240 & 8000 $\delta_\text{in}^{\star}$ & 320 $\delta_\text{in}^{\star}$ & 100 $\pi \delta_\text{in}^{\star}$ \\
M10C \cite{sciacovelli2021assessment}  & 10    & 350  & 3596 &  Adiab & 5520 & 256 & 240 & 8000 $\delta_\text{in}^{\star}$ & 320 $\delta_\text{in}^{\star}$ & 100 $\pi \delta_\text{in}^{\star}$  \\
M12 \cite{passiatore2022thermochemical}    & 12.48 & 594.3 & 4656 & $1800$  & 9660 & 480 & 512 & 3000 $\delta_\text{in}^{\star}$ & 120 $\delta_\text{in}^{\star}$ & 30 $\pi \delta_\text{in}^{\star}$  \\ [1ex]
\hline
\end{tabular}
\label{table:dnsparameters}
\end{table*}

\begin{figure}[!tb]
\centering
\includegraphics[width=0.49\textwidth]{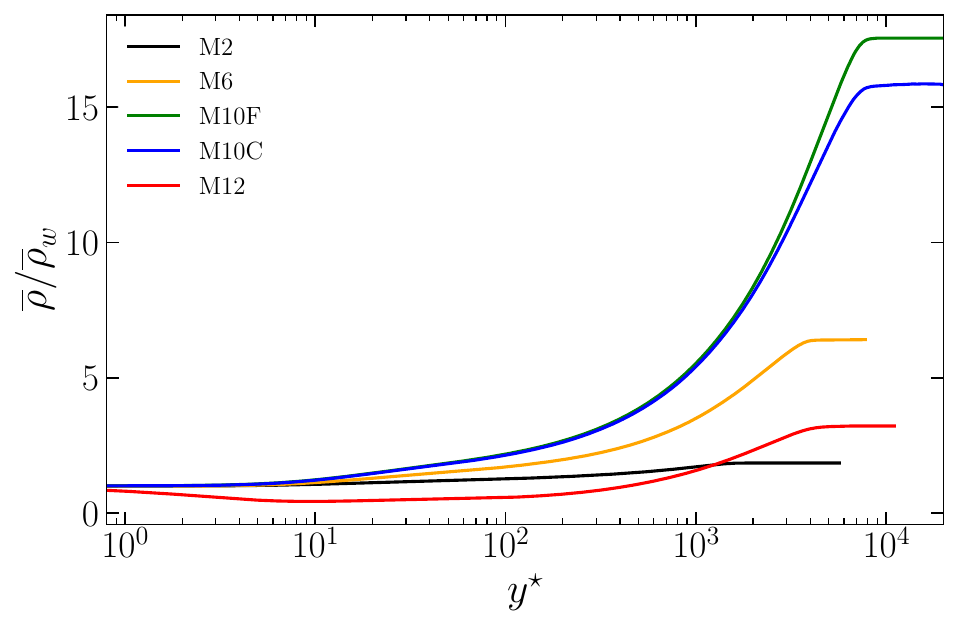}
\includegraphics[width=0.49\textwidth]{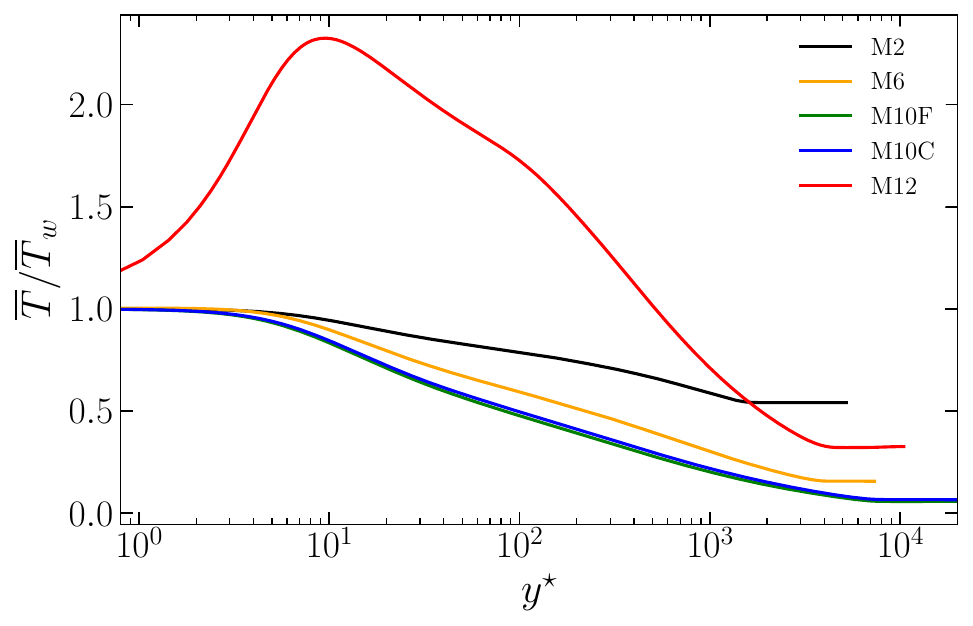}\\[-0.5cm]
\caption{Wall-normal profiles of density and temperature from the DNS database at the stations listed in table~\ref{table:dnsresults}.}
\label{fig:dns_data1}
\end{figure}

\begin{figure}[!tb]
\centering
\includegraphics[width=0.49\textwidth]{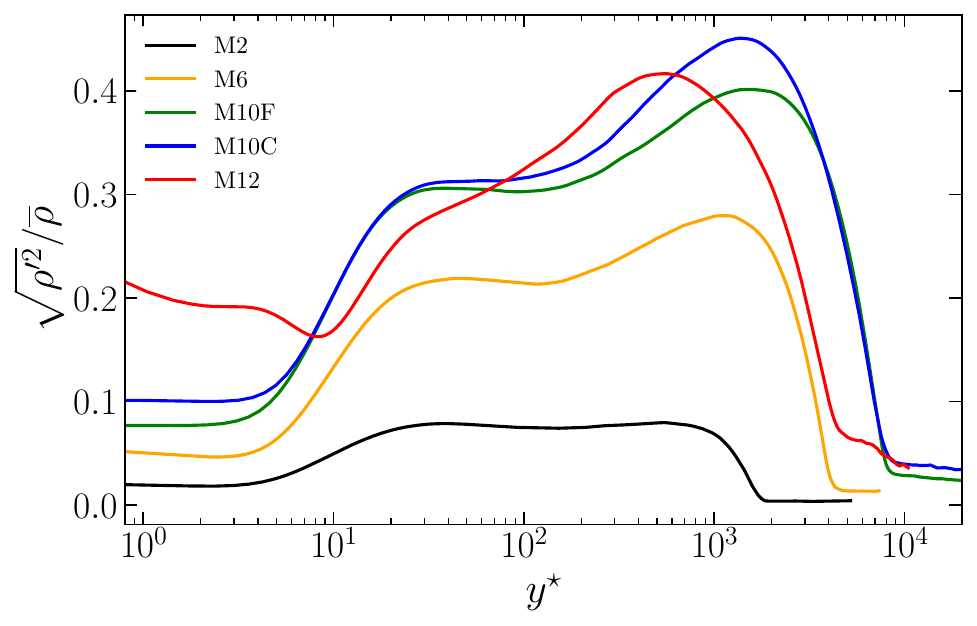}
\includegraphics[width=0.49\textwidth]{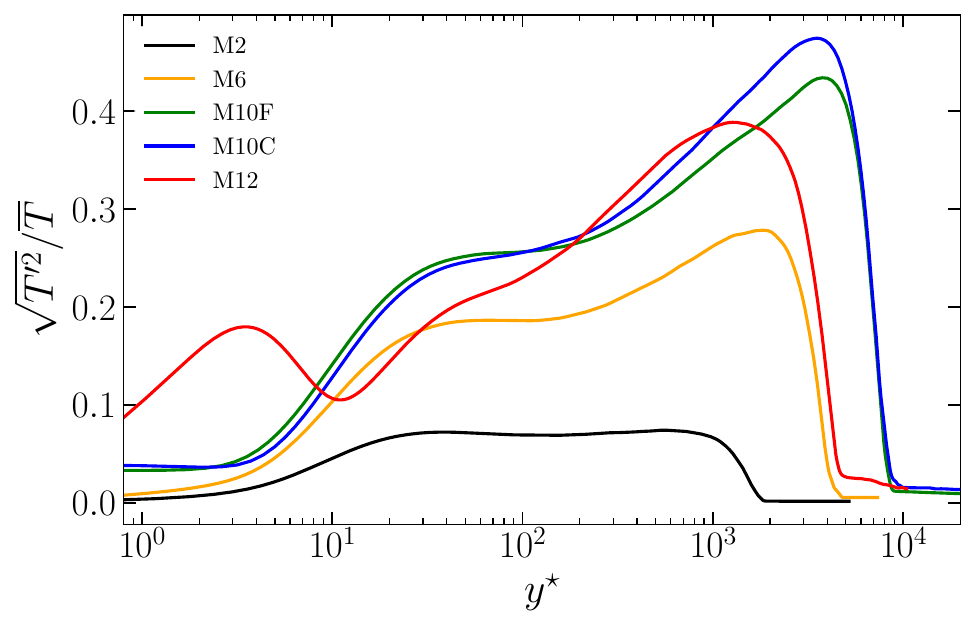}\\
\includegraphics[width=0.49\textwidth]{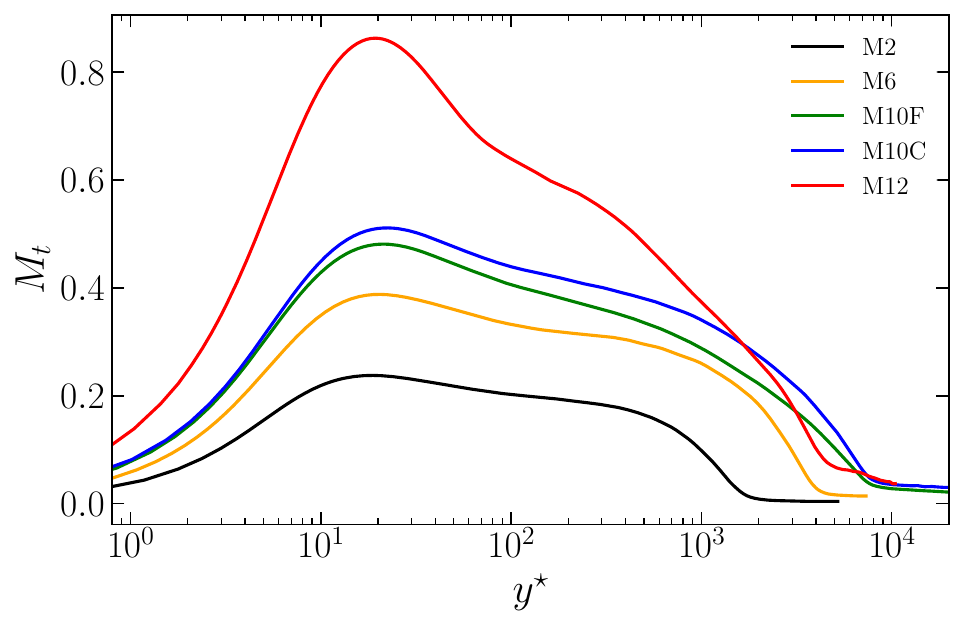}\\[-0.5cm]
\caption{Wall-normal profiles of normalized density fluctuations $\sqrt{\rho'^2}/\overline{\rho}$, normalized temperature fluctuations $\sqrt{T'^2}/\overline{T}$ and turbulent Mach number $M_t$ from the DNS database at the stations listed in table~\ref{table:dnsresults}.}
\label{fig:dns_data2}
\end{figure}

Five different DNS databases are considered in this work, spanning a wide range of thermodynamic conditions from the supersonic ($M_\infty = 2.25$) up to the hypersonic, high-enthalpy regime ($M_\infty = 12.48$) and with different levels of wall cooling. Details about the free-stream and boundary conditions, the number of grid points and the extent of the computational domains are given in table~\ref{table:dnsparameters}. Of note, the domain lengths for case M6 are normalized with respect to the boundary layer thickness at the outflow, $\delta_{99}^\text{out}$, since the inlet section corresponds to the leading edge of the flat plate. In the M2 and M6 runs, air is modeled as a single-species thermally- and calorically-perfect gas (see \cite{sciacovelli2020numerical}), whereas simulations denoted M10F, M10C and M12 make use of a thermally-perfect five-species mixture of N$_2$, O$_2$, NO, N and O. Non-catalytic wall conditions are considered for all the multi-species cases. Run M10C uses finite-rate chemistry models \cite{sciacovelli2021assessment}, whereas M10F is performed  by taking the M10C calculation and imposing a frozen-chemistry assumption (unpublished data set). Lastly, case M12 considers the contextual presence of vibrational and chemical nonequilibrium \cite{passiatore2022thermochemical}. All the test cases are spatially-evolving boundary layers initiated at a specific inlet Reynolds number. Transition to turbulence is triggered by means of suction and blowing strategy. The interested reader might refer to \cite{sciacovelli2020numerical,sciacovelli2021assessment,passiatore2022thermochemical} for details about the governing equations and the numerical schemes used in the simulations.
For a given variable $f$, in the following we denote with $\overline{f} = f - f'$ the standard time- and spanwise average, being $f'$ the corresponding fluctuation, whereas $\widetilde{f}= f - f''= \overline{\rho f} / \overline{\rho}$ denotes the density-weighted Favre averaging, with $f''$ the Favre fluctuation.
In order to carry out comparisons, profiles were extracted at streamwise stations corresponding to well-developed turbulent flow. Due to the very different flow conditions, the friction Reynolds numbers $Re_{\tau}= \overline{\rho}_w u_{\tau} \delta_{99}/\overline{\mu}_w$ (with $u_\tau = \sqrt{\overline{\tau}_w / \overline{\rho}_w}$) of the considered simulations do not match with each other. Nevertheless, profiles of the turbulent statistics were found to exhibit reasonable self-similarity at the chosen stations, thus implying that the Reynolds number effects do not play a significant role in the comparisons among the various cases. Some properties of the stations selected for the analysis are given in table~\ref{table:dnsresults}.\\
Figure~\ref{fig:dns_data1} displays wall-normal density and temperature profiles against the semi-locally scaled wall distance $y^\star$. All cases except M12 exhibit monotonic evolutions, with decreasing density and increasing temperature going from the free stream towards the flat plate wall. The similar trend is to be ascribed to the boundary condition imposed for the wall temperature, which is close (or equal to) the adiabatic value. The M12 run, on the contrary, features a strongly-cooled wall condition leading to a large friction heating inside the boundary layer; the maximum and minimum value for $\overline{T}/\overline{T}_w$ and $\overline{\rho}/\overline{\rho}_w$, respectively, are therefore achieved in the buffer layer. The normalized density fluctuations, temperature fluctuations and the turbulent Mach number $M_t = \sqrt{u_i'^2}/\overline{c}$ ($u_i$ denoting the velocity fluctuations and $c$ the speed of sound) are shown in the top, middle and bottom panel of Figure~\ref{fig:dns_data2}, respectively. In all cases, $M_t$ is shown to peak in the buffer layer and its values get larger with the increasing free-stream Mach number. The much larger $M_t$ registered for the M12 case is due to the presence of a significant wall cooling imposed by the isothermal wall condition, which is also responsible of the peak of temperature fluctuations in the inner region. Furthermore, the large near-wall temperature gradient leads in turn to increased density fluctuations towards the wall. The global maximum of the thermodynamic fluctuations is achieved close to the boundary layer edge, owing to the strong intermittency caused by the alternation of turbulent bulges and laminar-like flow regions. M10F and M10C runs display larger variations with respect to M12 because of the much smaller free-stream temperature. Since chemical dissociation reactions are of endothermic nature, the mean temperature values $\overline{T}$ for M10C tend to be smaller than those for M10F, leading to higher fluctuation levels for the former case.

\begin{table*}[!tb]
\centering
\caption{Data from the DNS at the streamwise stations selected for the analysis. Streamwise Reynolds number $Re_x$, friction Reynolds number $Re_\tau$, momentum-thickness-based Reynolds number $Re_\theta$, friction Mach number $M_\tau$, and grid resolutions in the streamwise ($\Delta x^+$), wall-normal ($\Delta y^+$) and spanwise ($\Delta z^+$) directions, respectively.}
\vspace{2mm}
\begin{tabular}{| c c c c c c c c |}
\hline
Case & $Re_x \cdot 10^{-6}$ & $Re_\tau$ & $Re_\theta$ & $M_\tau$ & $\Delta x^+$ & $\Delta y^+$ & $\Delta z^+$ \\[0.5ex]
\hline\hline
M2 \cite{sciacovelli2021assessment}    & 2.56 & 607   &  3538  & 0.075 & 4.4 & 0.59 & 2.7 \\
M6  \cite{sciacovelli2020numerical}    & 11.8 & 402   &  5972  & 0.122 & 3.8 & 0.26 & 2.1 \\
M10F                                   & 48.1 & 231   &  10875 & 0.154 & 3.7 & 0.42 & 3.4 \\
M10C \cite{sciacovelli2021assessment}  & 48.1 & 272   &  10411 & 0.157 & 4.5 & 0.50 & 4.1 \\
M12 \cite{passiatore2022thermochemical}& 15.2 & 1065  &  5491  & 0.213 & 7.6 & 0.67 & 4.5 \\
\hline
\end{tabular}
\label{table:dnsresults}
\end{table*}

\subsection{Velocity scalings}
In supersonic and hypersonic TBL flows, the mean velocity profiles deviate from the incompressible ones due to local variation in fluid properties. Wall friction converts kinetic energy into thermal energy, increasing the fluid temperature in the near wall region. This phenomenon, known as friction heating, leads to a decrease of density and an increase of dynamic viscosity, i.e. a lower local Reynolds number.
As a result, the viscous subregion in the velocity profiles thickens, and the slope of the logarithmic region is modified when plotted in the standard (incompressible) scalings, i.e. $y^+= \overline{\rho}_w u_{\tau} y/\overline{\mu}_w$ and $u^+ = \overline{u}/u_{\tau}$, which presume constant fluid properties.
\begin{table}[!tb]
\centering
\caption{List of transformations for the wall distance and the mean velocity as shown in equation~\eqref{eq:velscaling}, with $R=\frac{\overline{\rho}}{\overline{\rho}_w}$ and $M=\frac{\overline{\mu}}{\overline{\mu}_w}$.} \label{table:velscaling}
\vspace{0.2cm}
\renewcommand{\arraystretch}{1.25}
\begin{tabular}{lccc}
\hline
Transformation & Acronym & Wall distance $f_\text{I}$ & Mean velocity $g_\text{I}$\\
\hline
Van Driest \cite{van1951turbulent} & (VD) & 1& $R^{1/2}$\\
Trettel and Larsson \cite{trettel2016mean} & (TL) & $\frac{\partial}{\partial y} \Big( \frac{yR^{1/2}}{M} \Big)$ & $M \frac{\partial}{\partial y} \Big( \frac{yR^{1/2}}{M} \Big) $ \\[5pt]
Volpiani \cite{volpiani2020data} & (V) & $\frac{R^{1/2}}{M^{3/2}}$ & $\frac{R^{1/2}}{M^{1/2}}$ \\[5pt]
Griffin \emph{et al.} \cite{griffin2021velocity} & (G) & 1 & $\text{St}^+ \frac{\partial y^\star}{\partial y}$\\[5pt]
Hasan \emph{et al.} \cite{hasan2023incorporating} & (H) & 1 & $ \frac{1 + \kappa y^\star D^c}{1 + \kappa y^\star D^i} \Big( 1 - \frac{y}{\delta_v^\star} \frac{\text{d} \delta_v^\star}{\text{d} y} R^{1/2} \Big)$\\
\hline
\end{tabular}
\end{table}

Based on the Morkovin hypotesis \cite{morkovin1962effects}, multiple scalings have been developed with the objective of mapping the compressible mean flow profiles on the incompressible ones by taking into consideration the density and viscosity variations.
Considering the general transformations proposed by Modesti and Pirozzoli \cite{modesti2016reynolds}
\begin{equation} \label{eq:velscaling}
    y_\text{I} = \int_{0}^{y} f_\text{I} \, \text{d}y \qquad
    u_\text{I} = \int_{0}^{\tilde{u}} g_\text{I} \,\text{d}\widetilde{u}, \\
\end{equation}
where $f_\text{I}$, $g_\text{I}$ are mapping functions, usually depending on the ratios $R=\frac{\overline{\rho}}{\overline{\rho}_w}$ and $M=\frac{\overline{\mu}}{\overline{\mu}_w}$, and $y_\text{I}$, $u_\text{I}$ are the equivalent incompressible variables obtained by applying the transformation. A list of the considered transformations is shown in table~\ref{table:velscaling}. The first attempt at finding a suitable scaling was described by Van Driest \cite{van1951turbulent}, who only took into account wall-normal mean density variations. Its limitations (it is only valid for adiabatic walls) led to further developments in order to account for a wider range of condtions. Trettel and Larsson \cite{trettel2016mean} derived a velocity scaling by imposing a log-layer condition on the transformation and a constraint on the stress balance in the inner layer. Volpiani \textit{et al.} \cite{volpiani2020data} used
data-driven methods based on DNS datasets to calibrate the exponents for a simple power-law transformation. Due to the limited parameter space used for calibration, the model was shown not to work properly for channel and pipe flows. More recently, Griffin \textit{et al.} \cite{griffin2021velocity} proposed a total-stress-based transformation that does not depend on the thermal conditions at the wall. It can be derived from the stress scaling considering 
%
\begin{equation}
    \text{St}^+ = \frac{\tau^+ \text{S}^+_\text{eq}}{\tau^+ +\text{S}^+_\text{eq} -\text{S}^+_\text{TL} }, \qquad
    S^+_\text{eq} = \frac{1}{\mu^+} \frac{\partial U^+}{\partial  y^\star}, \qquad
    S^+_\text{TL} = \mu^+ \frac{\partial U^+}{\partial  y^+}, \qquad
    \tau^+ = \tau^+_{v} + \tau^+_{R}
\end{equation}
where $\text{St}^+$ is a generalised non dimensional mean stress, that behaves as $S^+_\text{TL}$ in the viscous sublayer and as $S^+_\text{eq}$ in the log-layer. The G transformation improved the results for compressible boundary layers; however, it is inaccurate for ideal gas flows with non-air-like viscosity laws and for supercritical flows \cite{bai2022compressible}. A newly derived transformation from Hasan \emph{et al.} \cite{hasan2023incorporating} builds upon the TL one, and take into account the intrinsic compressibility effects that are deemed to be the origin of the log-law shift often experienced for high-Ma number flows. Taking the friction Mach number $\text{Ma}_\tau$ as the most suitable parameter to quantify the effects, they propose the transformation shown in table~\ref{table:velscaling}, with
\begin{equation}
    D^i = \left[ 1 - \exp\left(-\frac{y^\star}{A^+}\right) \right]^2, \qquad D^c = \left[ 1 - \exp\left(-\frac{y^\star}{A^+ + f(M_\tau)}\right) \right]^2, \quad
\end{equation}
being $\kappa = 0.41$, $A^+=17$, $f(M_\tau) = 19.3 M_\tau$.\\
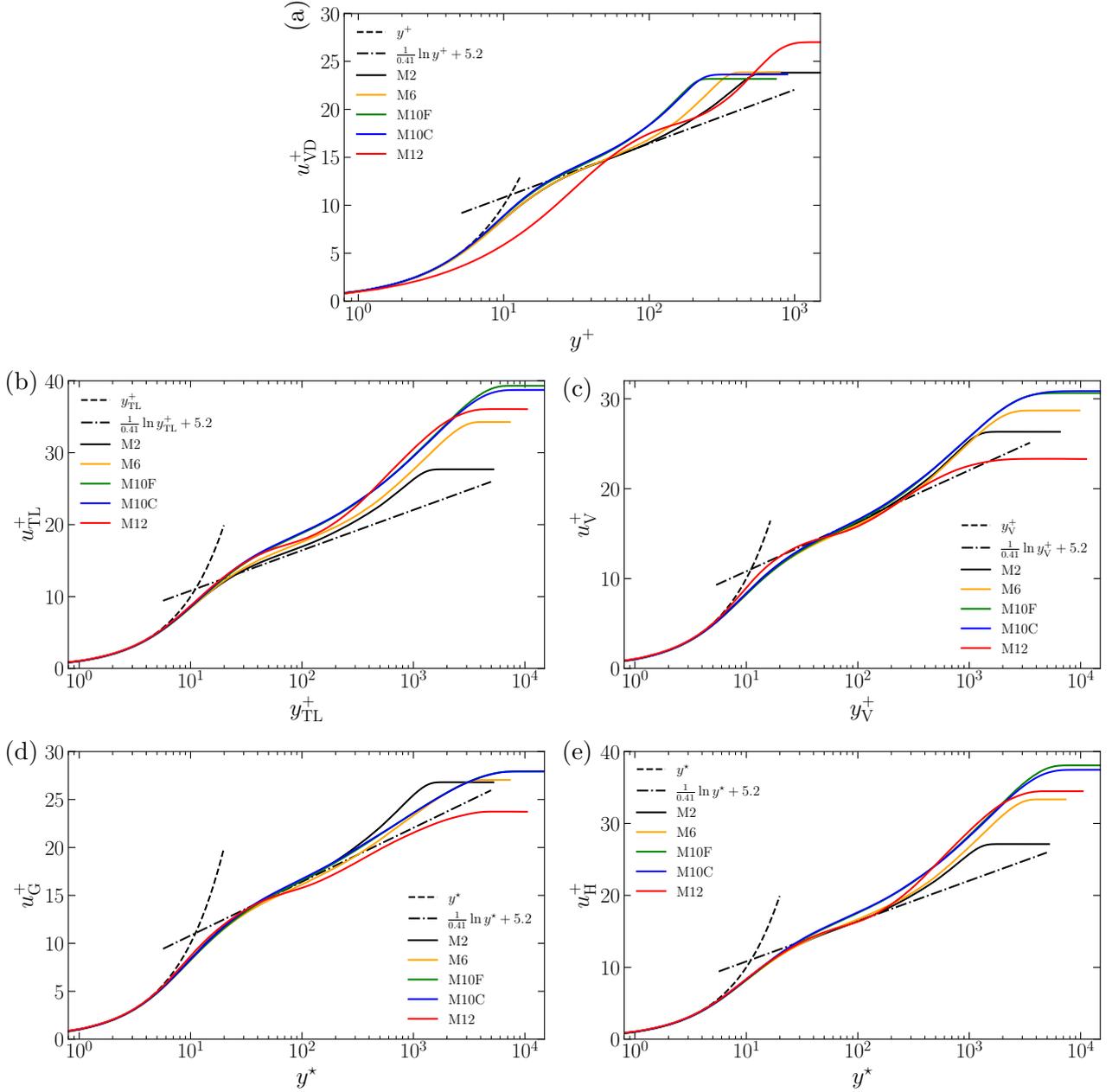
\begin{figure}[!tb]
\centering
 \begin{tikzpicture}
   \node[anchor=south west,inner sep=0] (a) at (0,0) {\includegraphics[width=0.49\textwidth]{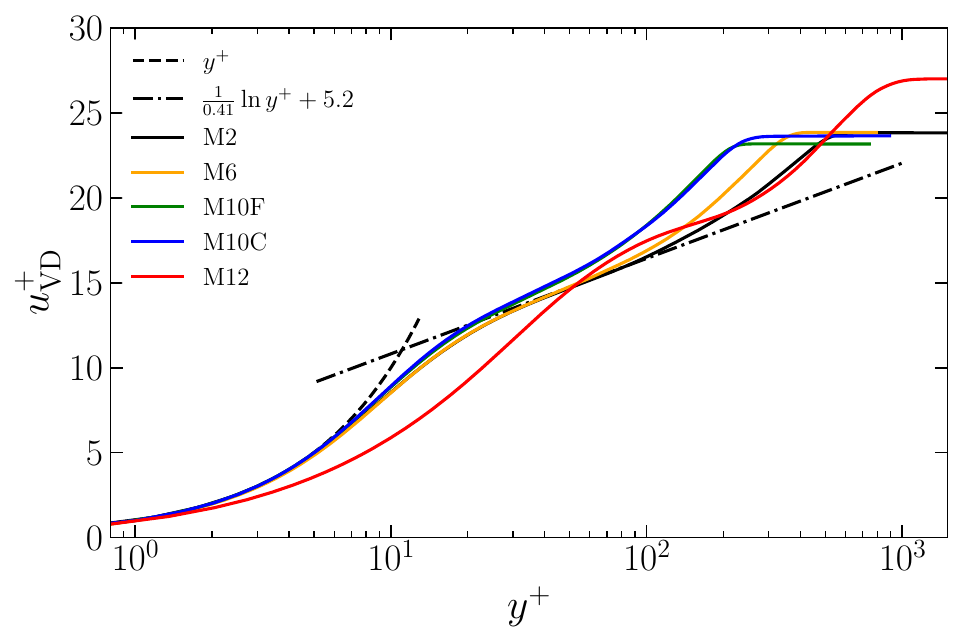}};
   \begin{scope}[x={(a.south east)},y={(a.north west)}]
     \node [align=center,rotate=90] at (0.03,0.95) {(a)};
   \end{scope}
 \end{tikzpicture}\\
\begin{tikzpicture}
   \node[anchor=south west,inner sep=0] (a) at (0,0) {\includegraphics[width=0.49\textwidth]{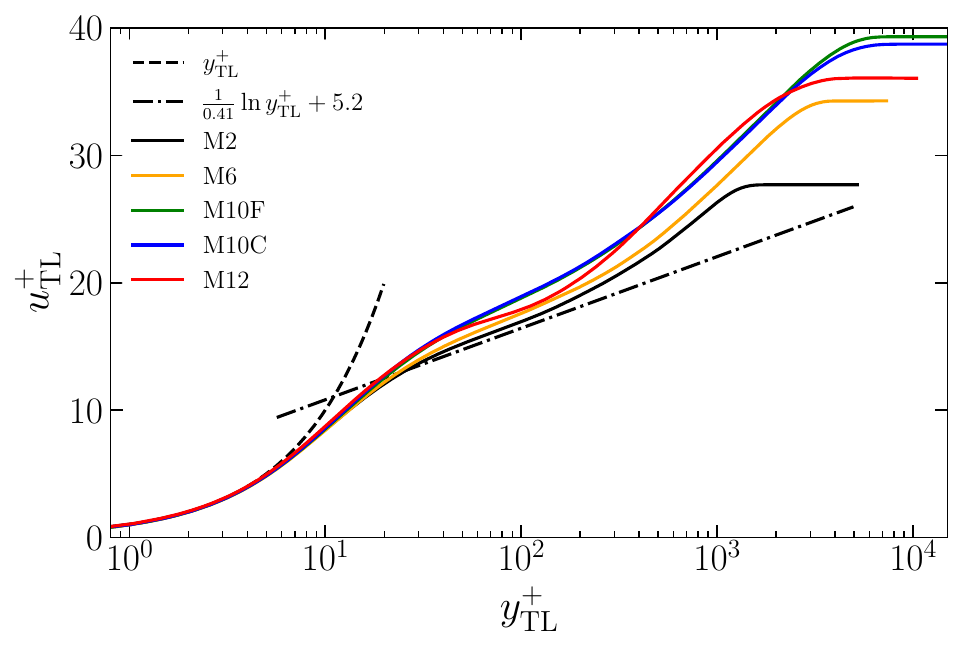}};
   \begin{scope}[x={(a.south east)},y={(a.north west)}]
     \node [align=center] at (0.03,0.95) {(b)};
   \end{scope}
 \end{tikzpicture}
 \hspace{-0.25cm}
 \begin{tikzpicture}
   \node[anchor=south west,inner sep=0] (a) at (0,0) {\includegraphics[width=0.49\textwidth]{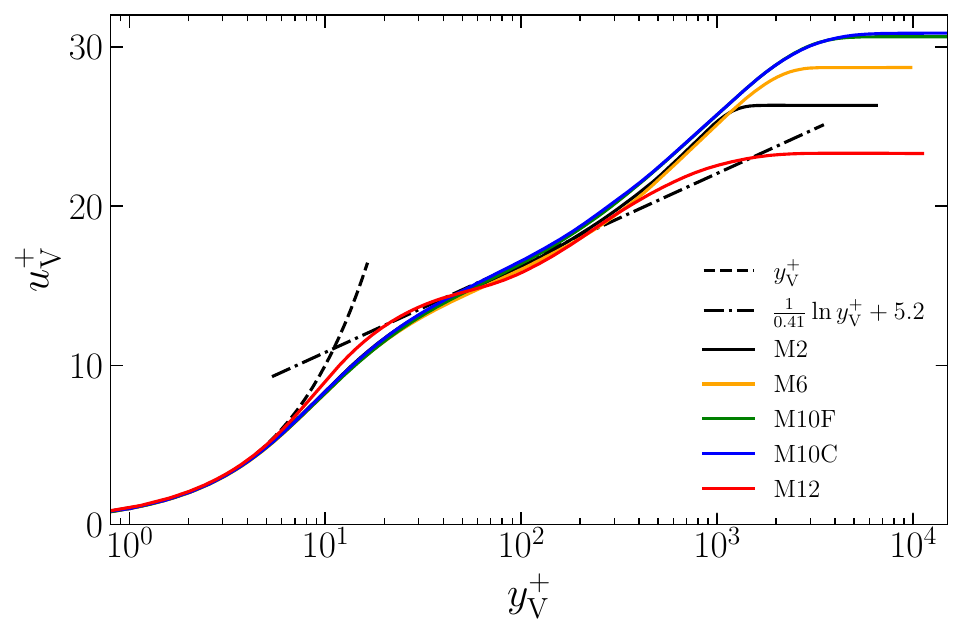}};
   \begin{scope}[x={(a.south east)},y={(a.north west)}]
     \node [align=center] at (0.03,0.95) {(c)};
   \end{scope}
 \end{tikzpicture}\\
 \begin{tikzpicture}
   \node[anchor=south west,inner sep=0] (a) at (0,0) {\includegraphics[width=0.49\textwidth]{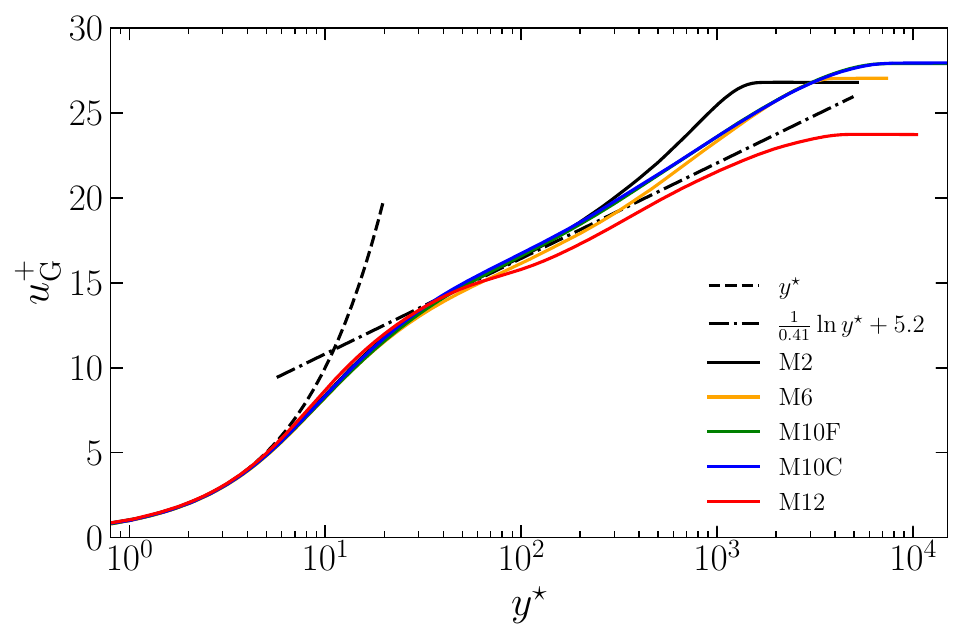}};
   \begin{scope}[x={(a.south east)},y={(a.north west)}]
     \node [align=center] at (0.03,0.95) {(d)};
   \end{scope}
 \end{tikzpicture}
 \hspace{-0.25cm}
 \begin{tikzpicture}
   \node[anchor=south west,inner sep=0] (a) at (0,0) {\includegraphics[width=0.49\textwidth]{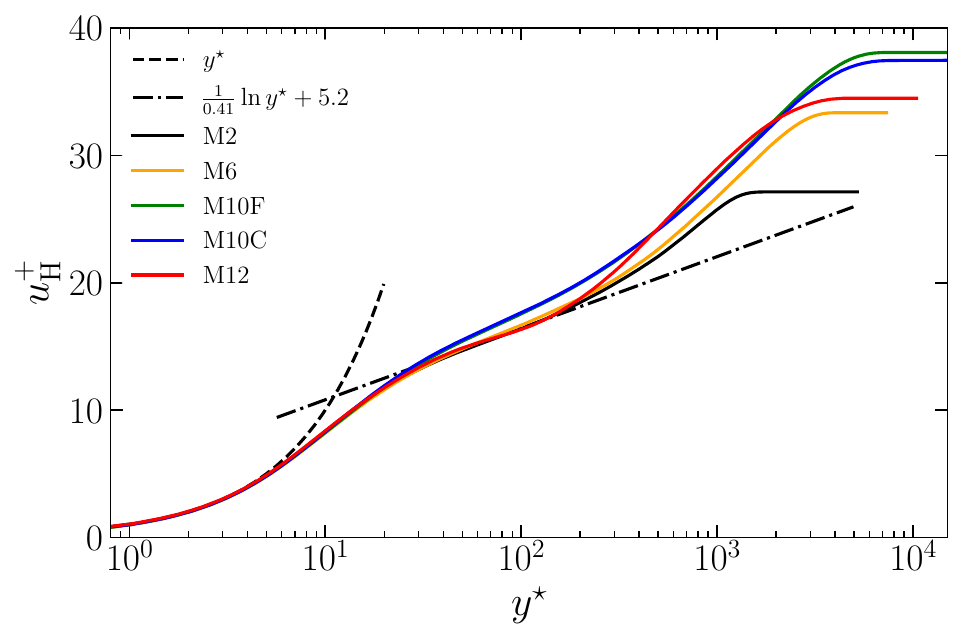}};
   \begin{scope}[x={(a.south east)},y={(a.north west)}]
     \node [align=center] at (0.03,0.95) {(e)};
   \end{scope}
 \end{tikzpicture}\\[-0.5cm]
\caption{Wall-normal profiles of the velocity profiles according to the transformation of van Driest (a), Trettel-Larsson (b), Volpiani (c), Griffin \emph{et al.} (d) and Hasan \emph{et al.} (e), from the DNS database at the stations listed in table~\ref{table:dnsresults}.}
\label{fig:Velocity_scalings}
\end{figure}
Figure \ref{fig:Velocity_scalings} shows the proposed transformations plotted against the classical incompressible law of the wall. The Van Driest scaling collapses the near wall region for all simulations except the M12; this trend is caused by the presence of a strong wall cooling which strongly deviates from the adiabatic wall assumption on which the model is derived. In the log region the model fails to collapse all the simulations with the analytical laws although a better matching can be found using a wall scaling.
The scaling of Volpiani offers excellent matching in the near wall region for all the cases, and retains good matching in the first part of the log region. However only the lower $M_{\infty}$ cases present a good matching in the log region. This can probably be related to the fact that the model coefficients have been calibrated using data from simulations up to $M_{\infty}=6$.
Considering the TL transformation, an excellent agreement is found in the near wall region, which was to be expected as the model was derived by imposing a stress-balance condition for the inner layer. A noticeable scattering is present however in the log region for the M12 simulation while the other cases match up to the outer region.
With regard to the G transformation, the near-wall region is matched perfectly as this model tends to the TL one in the inner layer; in the log region good matching is maintained for all the cases considered, since the transformation is able to account for the shear stress (large in this region).
Lastly, the scaling of Hassan \emph{et al.} provides also a satisfactory agreement in both the inner and outer layers; the velocity profiles for cases M10F and M10C are moved slightly upward with respect to the other ones, most likely because of the small Reynolds numbers at the selected station ($Re_\tau < 300$ for both cases).
Globally, the results show that the G and H transformations are able to provide more accurate results in a wide range of $M_{\infty}$ and wall temperature conditions.

\section{Equations and models}
\label{sec:models}
\subsection{Favre-averaged Navier--Stokes equations}
The basis of compressible turbulence models are the Favre-averaged compressible Navier--Stokes equations:
\begin{align}\label{eq:NS_mass}
    \frac{\partial \overline{\rho}}{\partial t} +  \frac{\partial (\overline{\rho }\widetilde{u_i})}{\partial x_i} & = 0 \\ \label{eq:NS_momentum}
    \frac{\partial \overline{\rho}\widetilde{u_i}}{\partial t} +  \frac{\partial \overline{\rho} \widetilde{u_i} \widetilde{u_j}}{\partial x_j} & = - \frac{\partial \overline{p}}{\partial x_i} + \frac{\partial}{\partial x_j}\left(\overline{\tau}_{ij} - \overline{\rho u_{i}''u_{j}''}\right)\\
    \notag
    \frac{\partial}{\partial t} \left[\overline{\rho}\widetilde{E} + \frac{\overline{\rho u_i''u_i''}}{2}\right] +
    \frac{\partial}{\partial x_j}\left[\overline{\rho} \widetilde{u_j}\left(\widetilde{H} + \frac{\overline{\rho u_i''u_i''}}{2}\right)\right] & = -\frac{\partial}{\partial x_j}\left[ \overline{q}_{j} +\overline{\rho u_j'' h''} \right] + \frac{\partial}{\partial x_j}\left[\overline{\tau_{ij}u_i''} - \overline{\rho u_j''\frac{1}{2}u_i''u_i''}\right] \\ & \quad + \frac{\partial}{\partial x_j}\left[\left(\overline{\tau}_{ij} - \overline{\rho u_i''u_j''}\right)\widetilde{u}_j\right] \label{eq:NS_energy}
\end{align}
where $t$ and $x_i$ denote the time and space coordinates, $u_i$ the velocity components, $p$ the pressure and $\tau_{ij}$ the viscous stress tensor; $E = e + \frac{1}{2}u_iu_i$ the specific total energy ($e$ being the specific internal energy), $H$ the total enthalpy and $q_j$ the heat flux. In the momentum and total energy equations \eqref{eq:NS_momentum}-\eqref{eq:NS_energy}, the Favre-averaged Reynolds stress tensor $\tau^R_{ij} = \overline{ \rho u_{i}''u_{j}''}$ appears; other unclosed terms are the turbulent kinetic energy $\overline{\rho}k = \frac{1}{2}\overline{\rho u_{i}'' u_{i}''}$, the turbulent heat flux $q^\text{t}_{j} = \overline{\rho u_j'' h''}$, the molecular diffusion $\overline{\tau_{ij}u_i''}$ and the turbulent transport of turbulent kinetic energy $\overline{\rho u_j''\frac{1}{2}u_i''u_i''}$.

When the operating conditions are such that the flow enters in a chemical non-equilibrium regime, air can no longer be considered as an homogeneous thermally and calorically-perfect gas. A mixture of five species is considered, namely, N$_2$, O$_2$, NO, N and O; system~\eqref{eq:NS_mass}-\eqref{eq:NS_energy} is therefore complemented with species conservation equations, here given in instantaneous form:
\begin{equation}
\frac{\partial \rho_{n}}{\partial t}+\frac{\partial\left(\rho_{n} u_{j}\right)}{\partial x_{j}} =-\frac{\partial \rho_{n} u_{n j}^{D}}{\partial x_{j}}+\dot{\omega}_{n}
\label{eq:species_eq}
\end{equation}
where $\rho_n = \rho Y_n$ is the partial density of the $n$-th species ($Y_n$ being mass fraction), $u_{nj}^D$ the $n$-th species diffusion velocity and $\dot{\omega}_n$ the chemical production rate. Using a Favre averaging we obtain:
\begin{equation}
\frac{\partial \bar{\rho} \widetilde{Y}_{n}}{\partial t}+\frac{\partial\left(\bar{\rho} \widetilde{Y}_{n} \widetilde{u}_{j}\right)}{\partial x_{j}}= - \frac{\partial \overline{\rho Y_{n} u_{nj}^D}}{\partial x_{j}} + \overline{\dot{\omega}}_{n}+\frac{\partial}{\partial x_{j}} \overline{\rho u_{j}^{\prime \prime} Y_{n}^{\prime \prime}},
\end{equation}
where the new unclosed term $\frac{\partial}{\partial x_{j}} \overline{\rho u_{j}^{\prime \prime} Y_{n}^{\prime \prime}}$ is the turbulent transport of chemical species.
The terms appearing from the averaging of the correction velocity term and the fluctuations of diffusive terms were found by Passiatore \textit{et al.} \cite{passiatore2021finite} to be negligible and therefore are not included.\\
In thermal non-equilibrium conditions, vibrational relaxation processes should also be considered. Following the two-temperature (2T) model of Park \cite{park1988two}, these processes are taken into account by adding a conservation equation for the vibrational energy $e_V$:
\begin{equation}
    \frac{\partial \rho e_{V}}{\partial t}+\frac{\partial \rho e_{V} u_{j}}{\partial x_{j}} = \frac{\partial}{\partial x_{j}}\left(-q_{V j}-\sum_{m=1}^{\mathrm{NM}} \rho_{m} u_{m j}^{D} e_{V m}\right) +\sum_{m=1}^{\mathrm{NM}}\left(Q_{\mathrm{TV} m}+\dot{\omega}_{m} e_{V m}\right). \label{eq:vibration_eq}
\end{equation}
Here, $\text{NM}$ is the number of molecular species ($\text{NM}=3$ in the current analysis), $q_{Vj}$ the vibrational contribution of the heat flux and $e_{V_m}$ the vibrational energy per unit of volume of the $m$-th species, given by
\begin{equation}
  e_{V m}=\frac{\theta_{m} R_{m}}{\exp \left(\theta_{m} / T_{V}\right)-1},
\end{equation}
$\theta_m$ being the characteristic vibrational temperature of each molecule, $R_m$ the gas constant and $T_V$ the vibrational temperature. Lastly, $Q_\text{TV}$ and $\dot{\omega}_m e_{Vm}$ are source terms denoting translational-vibrational energy exchanges and the vibrational energy variations due to chemical production/depletion, respectively. Additional information about the 2T model can be found in \cite{passiatore2022thermochemical}.
The Favre averaging of the vibrational energy equation \eqref{eq:vibration_eq} leads to:
\begin{equation}
\begin{aligned}
    \frac{\partial \bar{\rho} \widetilde{e}_{V}}{\partial t}+\frac{\partial}{\partial x_{j}} \left(\bar{\rho} \widetilde{e}_{V} \tilde{u}_{j}\right)&= \frac{\partial}{\partial x_{j}} \left(\overline{\rho u_{j}^{\prime \prime} e_{V}^{\prime \prime}}\right) + \frac{\partial}{\partial x_{j}} \left(-\overline{q_{V j}}+\sum_{m=1}^{\mathrm{NM}} \overline{\rho D_{m} \frac{\partial Y_{m}}{\partial x_{j}} e_{V m}}\right)\\
    &+\sum_{m=1}^{\mathrm{NM}}\left(\overline{{Q_{\mathrm{TV} m}}} + \overline{\dot{\omega}_{m}e_{V m}} \right),
    \label{rans_vibrational}
\end{aligned}
\end{equation}
where $\frac{\partial}{\partial x_{j}} \left(\overline{\rho u_{j}^{\prime \prime} e_{V}^{\prime \prime}}\right)$ is the turbulent transport term. Favre-averaging of equations \eqref{eq:species_eq} and \eqref{eq:vibration_eq} leads not only to turbulent transport of species and vibrational energy terms to be closed, but also to the averaged source terms that, due to the non-linearity of their expressions, need an appropriate closure model. These aspects are discussed in Sections~\ref{sec:mass} and \ref{sec:vibrational}.

\subsection{Turbulence closure models}
Throughout this work we solely focus on the so-called eddy-viscosity models, largely employed in engineering applications. These models rely on the Boussinesq approximation to express the constitutive law for the Reynolds stresses:
\begin{equation}
    \overline{\rho}\tau^R_{ij} = 2 \mu_t \left( \overline{S}_{ij} - \frac{1}{3}\overline{S}_{kk} \delta_{ij}\right) - \frac{2}{3} \overline{\rho}k \delta_{ij}
    \label{eq:stressre}
\end{equation}
where $\mu_t$ is the eddy viscosity, $\overline{S}_{ij}$ the mean strain-rate tensor and $\delta_{ij}$ the Kronecker symbol. Such an approximation reduces the modeling problem to the calculation of a single scalar, $\mu_t$. For two-equations models, the eddy viscosity is expressed as a function of two turbulent variables allowing to determine a turbulent length and velocity scale. For each of them a transport equation is written in order to calculate the eddy viscosity.
The first variable is generally the turbulent kinetic energy $k$, for which an exact transport equation can be derived from the momentum equation (see section~\ref{sec:tke}). Hereafter, we will focus on the SST $k$-$\omega$ model \cite{menter1992improved}, whose equations are presented in the next section.
The turbulent heat fluxes are generally modelled through a ``turbulent Fourier law'':
\begin{equation}
      q_{j,t} = -\frac{\mu_t c_p}{\text{Pr}_t}\frac{\partial \widetilde{T}}{\partial x_j}
       \label{eq:turbfourier}
\end{equation}
where $\text{Pr}_t$ is the turbulent Prandtl number, most often assumed to be constant and equal to 0.9 for turbulent air flows. In practice, this parameter is case-dependent and varies throughout the flow.
In \cite{passiatore2022thermochemical}, a ``vibrational Fourier law'' is proposed to model the corresponding term in the vibrational energy equation.
Similarly, turbulent mass fluxes arising in the Favre-averaged species transport are modelled through a turbulent Fick law by introducing again a, supposedly constant, turbulent Schmidt number $Sc_t$.
For passive scalar transport, the turbulent Schmidt and Prandtl number coincide, so that a common approximation consists in setting $Sc_{t,n}\approx 0.9$ for all species.

\section{A Priori Tests}\label{sec:results}
The validity of the above-mentioned assumptions, as well as supplementary assumptions used to close the turbulent kinetic energy transport equation are assessed against the exact unclosed terms extracted from the DNS statistics.
\subsection{Eddy viscosity and Reynolds stresses}
\begin{figure}[!tb]
\centering
\includegraphics[width=0.49\textwidth]{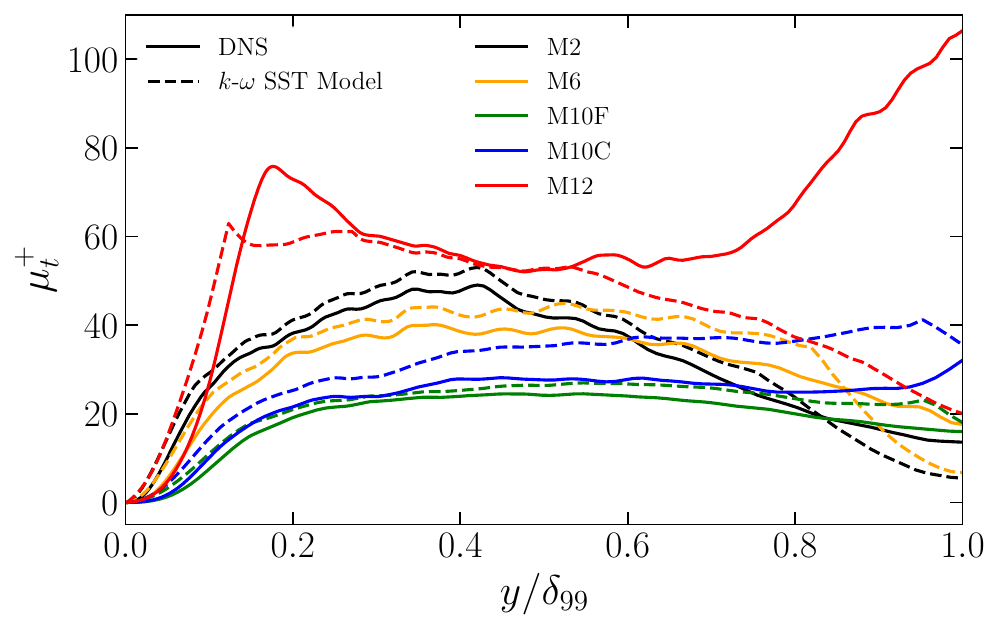}
\includegraphics[width=0.49\textwidth]{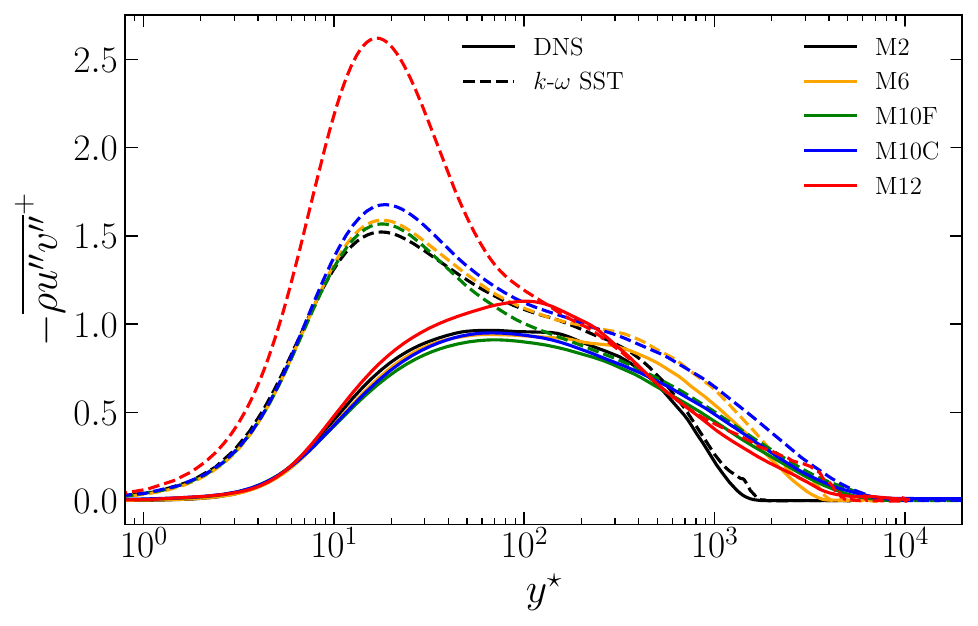}\\
\includegraphics[width=0.49\textwidth]{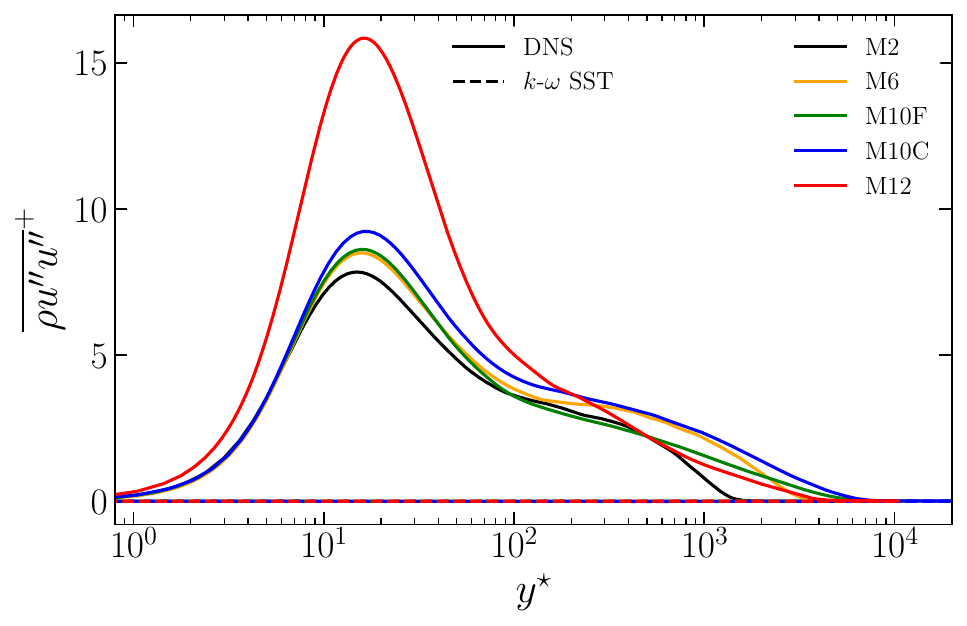}
\includegraphics[width=0.49\textwidth]{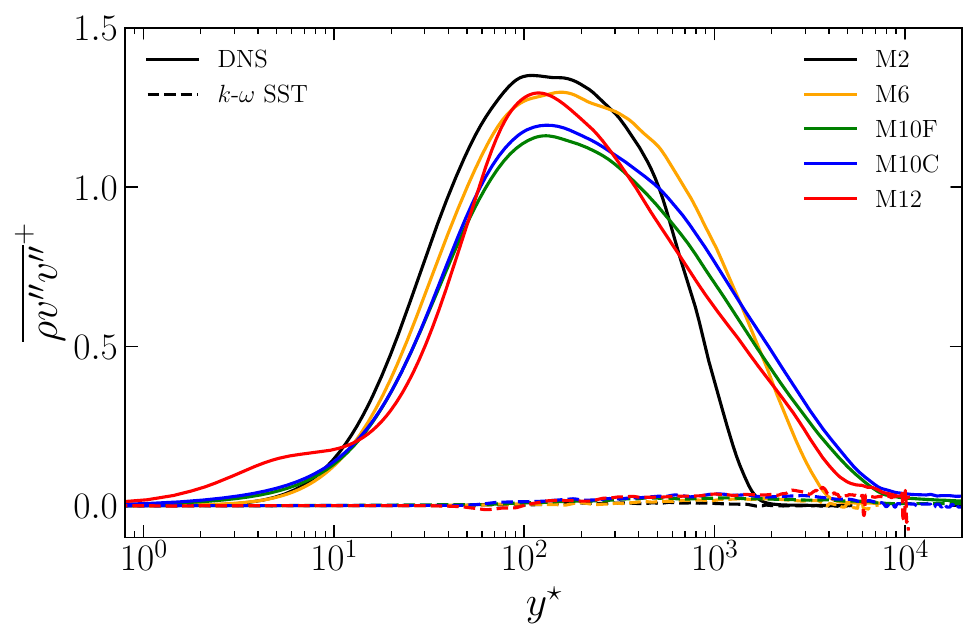}\\[-0.5cm]
\caption{Exact $\mu_t^+$ and Reynolds stresses from DNS versus their modelled counterparts.}
\label{fig:exact_reynolds_stresses}
\end{figure}

First, we discuss the validity of linear constitutive laws adopted for the Reynolds stresses, turbulent heat fluxes, and  mass fluxes.
The constitutive relation adopted for the Reynolds stresses is a linear eddy viscosity model (LEVM), of the form $\mu_t = \rho C_{\mu} \ell_t v_t$, where $\ell_t$ and $v_t$ are the characteristic length and velocity scales of turbulent structures, respectively, and $C_{\mu}$ a model constant to be defined. The model transport equations for the SST $k$-$\omega$ model in conservative form are
\begin{align}
\frac{\partial \overline{\rho} k }{\partial t} + \frac{\partial \overline{\rho} k \widetilde{u}_j}{\partial x_j} & = \tau^R_{ij} \frac{\partial \widetilde{u}_i }{\partial x_j} - \beta^*\overline{\rho}\omega k + \frac{\partial}{\partial x_j}\left[(\overline{\mu} + \sigma_k \mu_t)\frac{\partial k }{\partial x_j}\right]\\
\frac{\partial \overline{\rho} \omega}{\partial t}+\frac{\partial \overline{\rho} \omega \widetilde{u}_j}{\partial x_{j}} & = \frac{\overline{\rho}\gamma}{\mu_t} \tau^R_{ij} \frac{\partial \widetilde{u}_i }{\partial x_j} - \beta \overline{\rho} \omega^{2} +\frac{\partial}{\partial x_{j}}\left[\left(\overline{\mu}+\sigma_{\omega} \mu_t\right) \frac{\partial \omega}{\partial x_{j}}\right] +2\left(1-F_{1}\right) \frac{\overline{\rho}\sigma_{\omega 2} }{\omega} \frac{\partial k}{\partial x_{j}} \frac{\partial \omega}{\partial x_{j}}
\end{align}
with
\begin{equation}\label{eq:mutkappaomega}
     \mu_{t}=\frac{\overline{\rho} a_{1} k}{\max \left(a_{1} \omega, S F_{2}\right)}
\end{equation}
The values of the model constants are computed as $\phi = F_1 \phi_1 + (1-F_1) \phi_2$, where $\phi_1$ and $\phi_2$ denote the values for the original $k$-$\omega$ and the transformed $k$-$\varepsilon$ models, respectively. The constants of sets 1 and 2 are:
\begin{equation*}
    \sigma_{k1} = 0.85, \qquad \sigma_{\omega1} = 0.500, \qquad \beta_1 = 0.0750, \qquad \gamma_1 = \frac{\beta_1}{\beta^\star} - \frac{\sigma_{\omega1} \kappa^2}{\sqrt{\beta^\star}}
\end{equation*}
\begin{equation*}
    \sigma_{k2} = 1.00, \qquad \sigma_{\omega2} = 0.856, \qquad \beta_2 = 0.0828, \qquad \gamma_2 = \frac{\beta_2}{\beta^\star} - \frac{\sigma_{\omega2} \kappa^2}{\sqrt{\beta^\star}}
\end{equation*}
with $\beta^* = 0.09$, $\kappa = 0.41$, $a_1 = 0.31$. Moreover,
\begin{align*}
    F_1 &= \tanh (\arg_1^4) \qquad \text{with } \arg_1 = \min\left[ \max\left( \frac{\sqrt{k}}{\beta^\star\omega y}; \frac{500 \overline{\mu}}{y^2\overline{\rho}\omega} \right); \frac{4 \rho \sigma_{\omega2}k}{\text{CD}_{k\omega}y^2} \right]\\
    F_2 &= \tanh(\arg_2^2) \qquad \text{with } \arg_2 = \max \left( \frac{2 \sqrt{k}}{\beta^\star\omega y} ; \frac{500 \overline{\mu}}{y^2\overline{\rho}\omega} \right)
\end{align*}
with $\displaystyle \text{CD}_{k\omega} = \max \Big( 2 \rho \sigma_{\omega2} \frac{1}{\omega} \frac{\partial k}{\partial x_j} \frac{\partial \omega}{\partial x_j} , 10^{-20} \Big)$.
In figure~\ref{fig:exact_reynolds_stresses} we report the ``exact'' eddy viscosity coefficient, computed as: $\mu_t = -\overline{\rho u''v''}\frac{\partial \widetilde{u}}{\partial y}^{-1}$ and the modelled $\mu_t$ of equation~\eqref{eq:mutkappaomega} computed from the DNS data.
where we use the $k$ and $\omega$ resulting from the DNS statistics.
For the simple sheared flow at stake, the shear stress dominates the wall normal components, and the LEVM eddy viscosity (top panel) follows reasonably well the exact one for cases M2, M6 and also for the frozen-chemistry case M10F, indicating the effects of increasing friction heating are well-captured by the model overall. More significant deviations are observed for the finite-rate chemistry, adiabatic wall case M10C, for which the SST-model overpredicts the eddy viscosity significantly.
For such a case, the endothermic chemical reactions drain part of the energy supplied by the mean field, leading to reduced turbulent kinetic energy compared to the corresponding frozen case \cite{passiatore2021finite} and altering to some extent the near-wall characteristic time scales.
The largest discrepancies are obtained for the M12 case, for which thermal non-equilibrium effects combine with severe wall cooling. For this case, not only the near-wall profiles differ, but even larger deviations are observed at the boundary layer edge. In this region, turbulent velocity fluctuations vanish, but still fluctuating stresses persist as an effect of the strong acoustic radiation outside the boundary layer.
Note that since $k$ and $\omega$ are  obtained directly from DNS data, the above-mentioned discrepancies can be attributed to the expression used for $\mu_t$, and specifically to the empirical function $F_2$. In contrast, as expected for a LEVM, the normal components reconstructed from the Boussinesq model are severely underestimated, since they are  based only on the isotropic $- \frac{2}{3} \overline{\rho}k \delta_{ij}$ term.

\subsection{Turbulent Kinetic Energy equation}
\label{sec:tke}
Next, we assess the validity of models used for the closure of the turbulent kinetic energy equation. In the compressible case, this takes the form:
%
%
\begin{equation}\label{eq:exactbudget}
    \frac{\partial \overline{\rho} k}{\partial t} = -C + P + T_d + T_p + D_v + D_d + M + \Pi_d
\end{equation}
with
\begin{equation}
    C = \frac{\partial}{\partial x_j} (\overline{\rho} \widetilde{u}_j k), \qquad P = -\overline{\rho}\widetilde{u_i''u_j''}\frac{\partial \widetilde{u}_i}{\partial x_j}
\end{equation}
\begin{equation}
    T_d = -\frac{\partial}{\partial x_j} \Big(\frac{1}{2} \overline{\rho u_i''u_i''u_j''} \Big), \qquad T_p = -\frac{\partial}{\partial x_j}(\overline{p'u_j'})
\end{equation}
\begin{equation}
    D_v = \frac{\partial}{\partial x_j}(\overline{\tau_{ij}'u_i'}), \quad
    D_d = -\overline{\tau_{ij}'\frac{\partial u_i'}{\partial x_j}}, \quad
    M = \overline{u_i''}\left( \frac{\partial \overline{\tau_{ij}}}{\partial x_j} - \frac{\partial \overline{p}}{\partial x_i} \right), \quad
    \Pi_d = \overline{p' \frac{\partial u_i'}{\partial x_i}}
\end{equation}
where $C$, $P$, $T_d$, $T_p$, $D_v$, $D_d$, $M$ and $\Pi_d$ denote the contributions from convection, production, turbulent diffusion, velocity-pressure interaction, viscous diffusion, energy dissipation, mass flux contribution associated with density fluctuations and pressure dilatation, respectively.

\begin{figure}[!tb]
\centering
\includegraphics[width=0.49\textwidth]{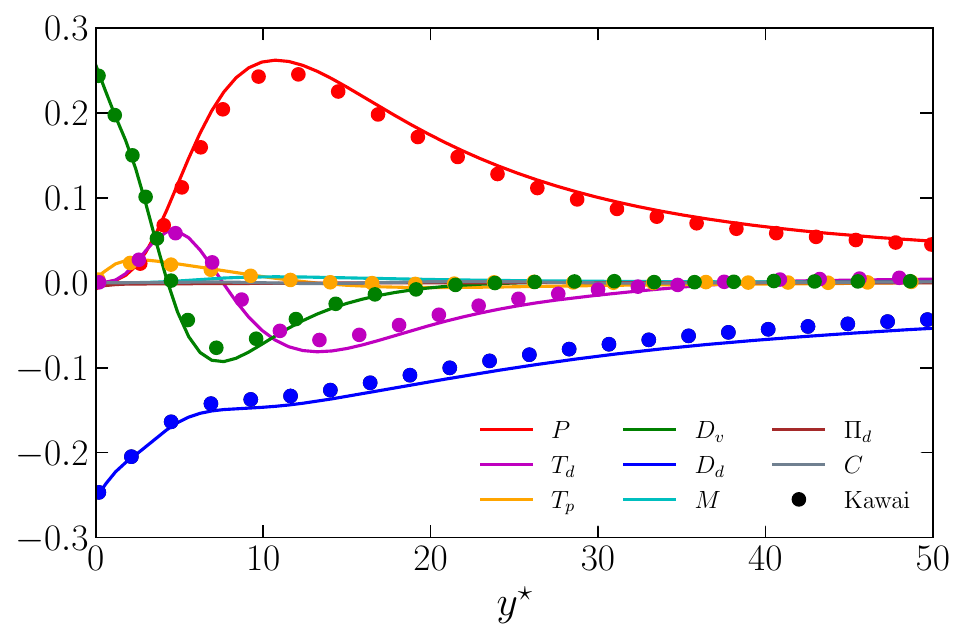}
\includegraphics[width=0.49\textwidth]{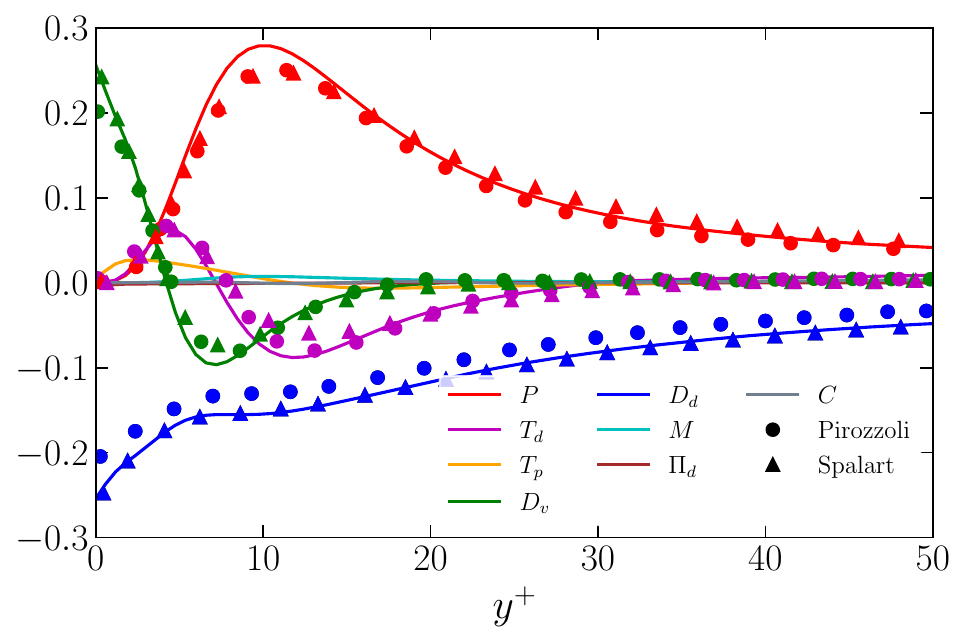}\\[-0.5cm]
\caption{Turbulent kinetic energy budget for case M2, compared to references data from Kawai~\cite{kawai2019heated}, Pirozzoli \emph{et al.}~\cite{pirozzoli2004directa} and Spalart~\cite{spalart1988direct}. Profiles on the left panel are scaled by $\overline{\mu}/\overline{\tau}_w^2$ and shown as a function of $y^\star$, whereas on the right panel they are scaled by $\overline{\mu}_w/\overline{\tau}_w^2$ and shown as a function of $y^+$.}
\label{fig:tke_budget_comp}
\end{figure}

\begin{figure}[!tb]
\centering
\includegraphics[width=0.8\textwidth]{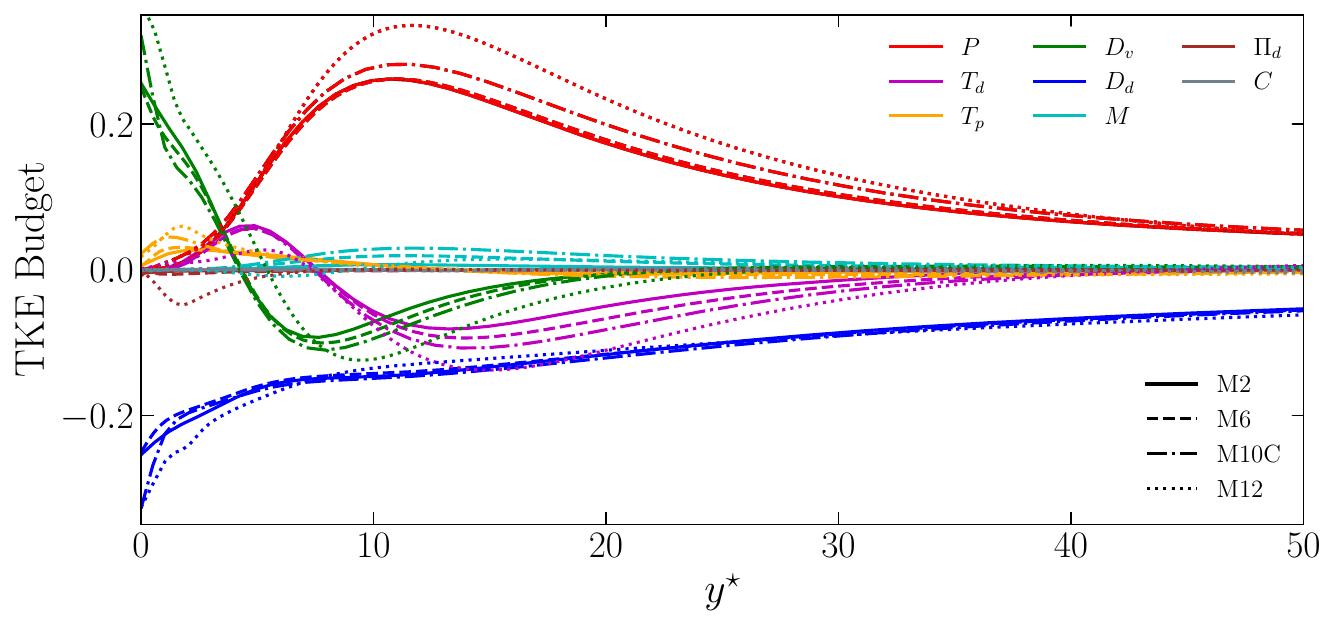}\\[-0.5cm]
\caption{Wall-normal profiles of the turbulent kinetic energy budgets for cases M2, M6, M10C and M12, shown in semi-local scaling.}
\label{fig:TKE_budget}
\end{figure}

The DNS database is first validated by comparison with data from the literature. Figure~\ref{fig:tke_budget_comp} shows the kinetic energy budget for case M2 compared with data from Kawai~\cite{kawai2019heated}, Pirozzoli \emph{et al.}~\cite{pirozzoli2004directa} and Spalart~\cite{spalart1988direct}. A very good agreement is registered for both the compressible and incompressible scalings; the minor discrepancies observed in the production and dissipation terms are mainly due to Reynolds number effects. The turbulent kinetic energy budgets for cases M2, M6, M10C and M12 are shown in figure~\ref{fig:TKE_budget} in semilocal scaling. Of note, the M10F profiles are superposed to the M10C and therefore are not shown. While the M2 and M6 cases share similar budgets, starting from M6 the magnitude of each budget term is shown to slightly increase with the free stream Mach number. The largest deviations are registered close to the wall for dissipation and viscous diffusion, and in the buffer layer for production and turbulent diffusion, respectively. Owing to the strong wall cooling, the pressure-dilation term for M12 is shown to be much larger with respect to the other cases. Overall, the pressure transport and pressure dilatation terms remain significantly smaller than the other terms even for the highest Mach number considered in this study.

Several unclosed terms appear in equation~\eqref{eq:exactbudget}, some of which are counterparts of those already present in the incompressible $k$ formulation. Some other, related to compressibility effects, also need to be modelled. In the following, we perform \emph{a priori} tests of various models existing in the literature. We restrict our attention to the terms contributing the most to the turbulent kinetic energy budget. Hereafter, all the terms are made non-dimensional by $\overline{\mu}/\overline{\tau}_w^2$ since taking into account $\overline{\mu}$ instead of $\overline{\mu}_w$ allows for better scaling.
\subsubsection{Turbulent Diffusion}
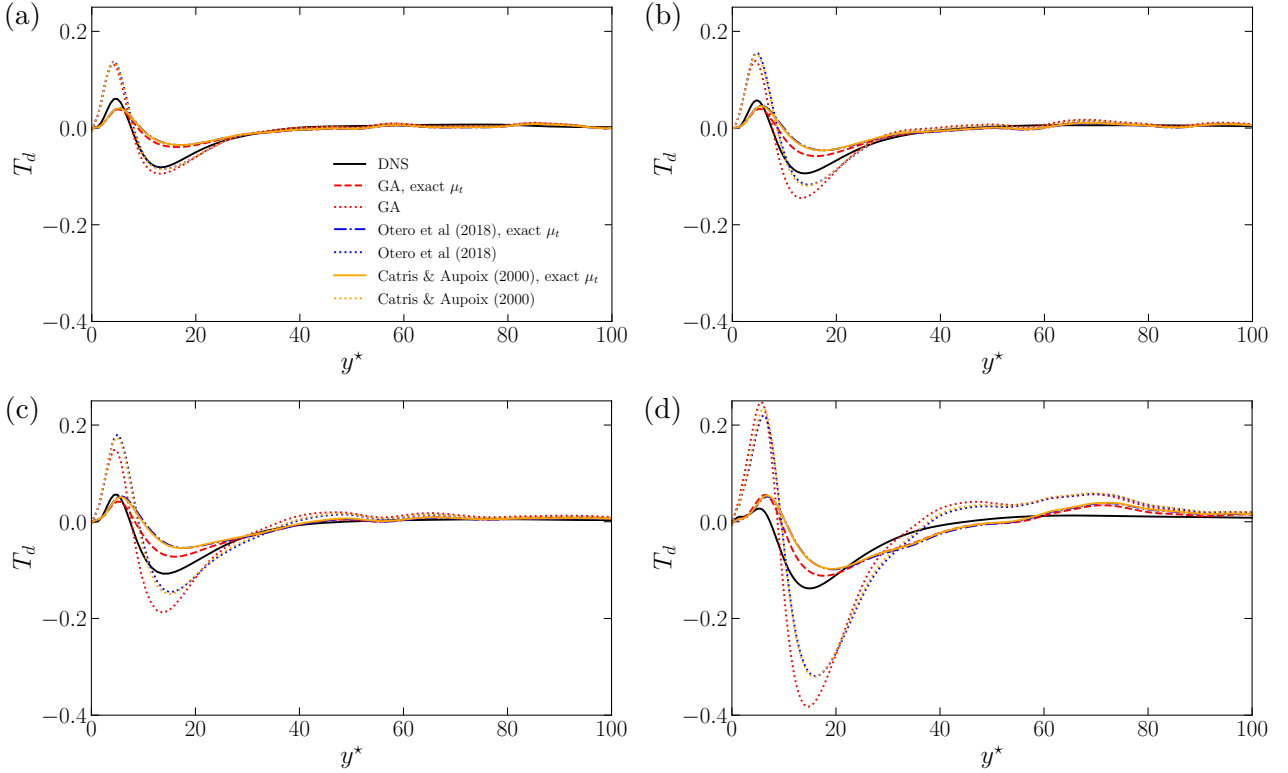
\begin{figure}[!tb]
\centering
\begin{tikzpicture}
   \node[anchor=south west,inner sep=0] (a) at (0,0) {\includegraphics[width=0.49\textwidth]{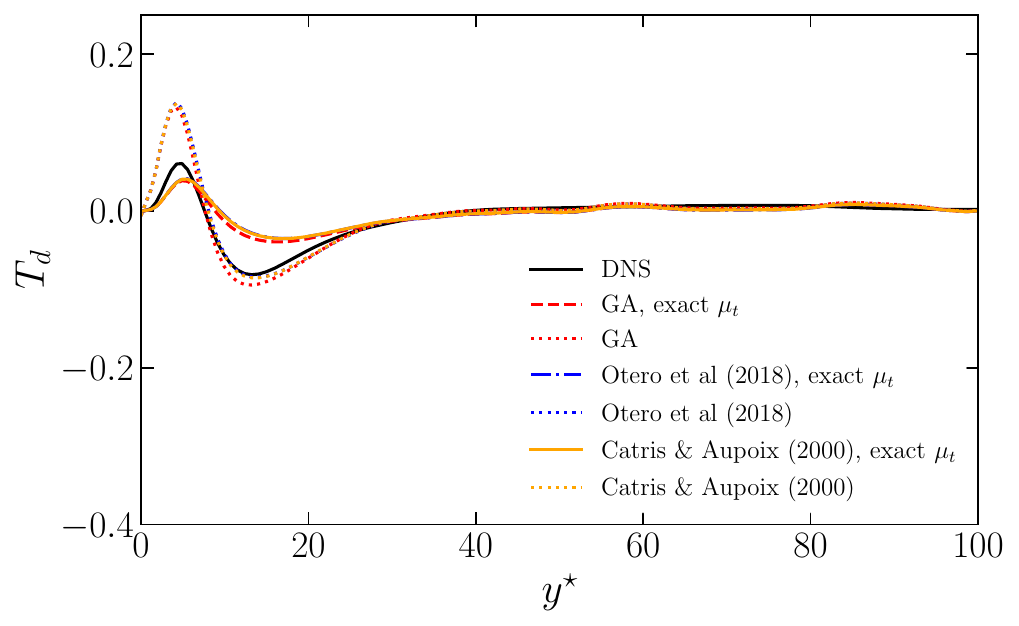}};
   \begin{scope}[x={(a.south east)},y={(a.north west)}]
     \node [align=center] at (0.03,0.95) {(a)};
   \end{scope}
 \end{tikzpicture}
 \hspace{-0.3cm}
 \begin{tikzpicture}
   \node[anchor=south west,inner sep=0] (a) at (0,0) {\includegraphics[width=0.49\textwidth]{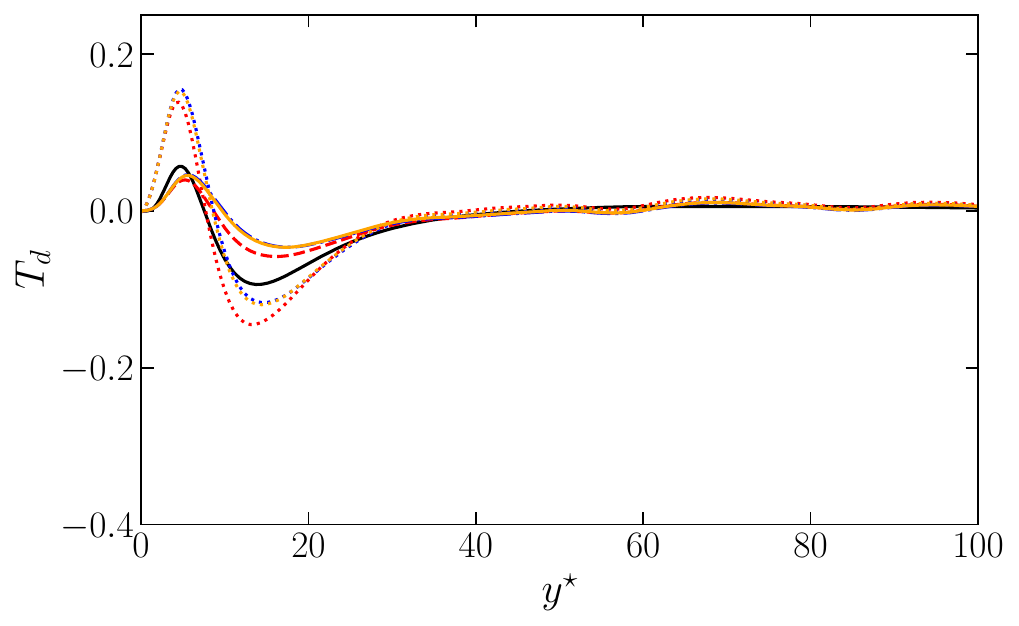}};
   \begin{scope}[x={(a.south east)},y={(a.north west)}]
     \node [align=center] at (0.03,0.95) {(b)};
   \end{scope}
 \end{tikzpicture}\\
 \begin{tikzpicture}
   \node[anchor=south west,inner sep=0] (a) at (0,0) {\includegraphics[width=0.49\textwidth]{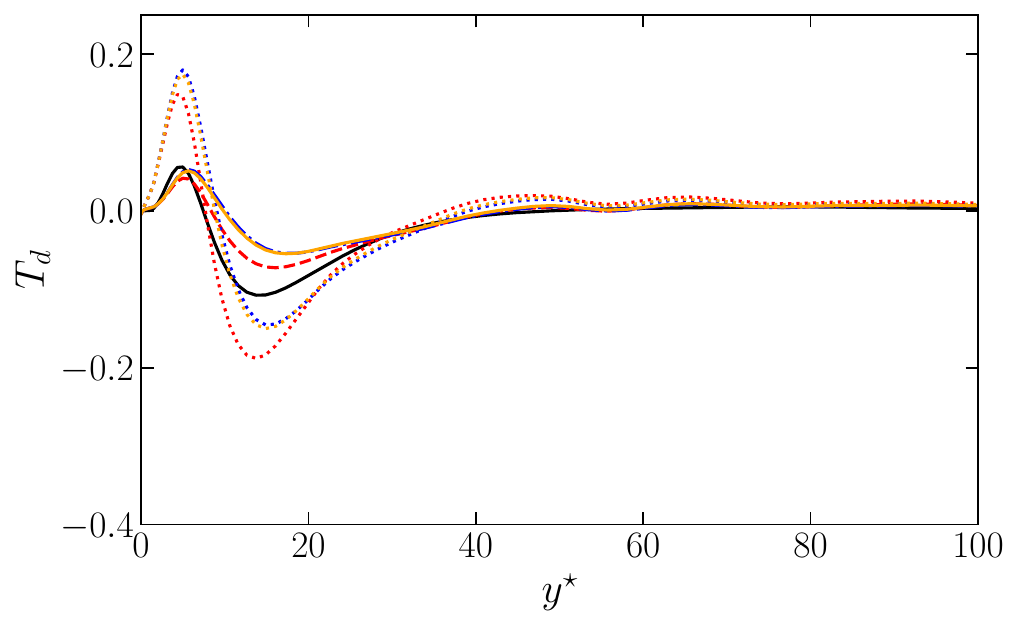}};
   \begin{scope}[x={(a.south east)},y={(a.north west)}]
     \node [align=center] at (0.03,0.95) {(c)};
   \end{scope}
 \end{tikzpicture}
 \hspace{-0.3cm}
 \begin{tikzpicture}
   \node[anchor=south west,inner sep=0] (a) at (0,0) {\includegraphics[width=0.49\textwidth]{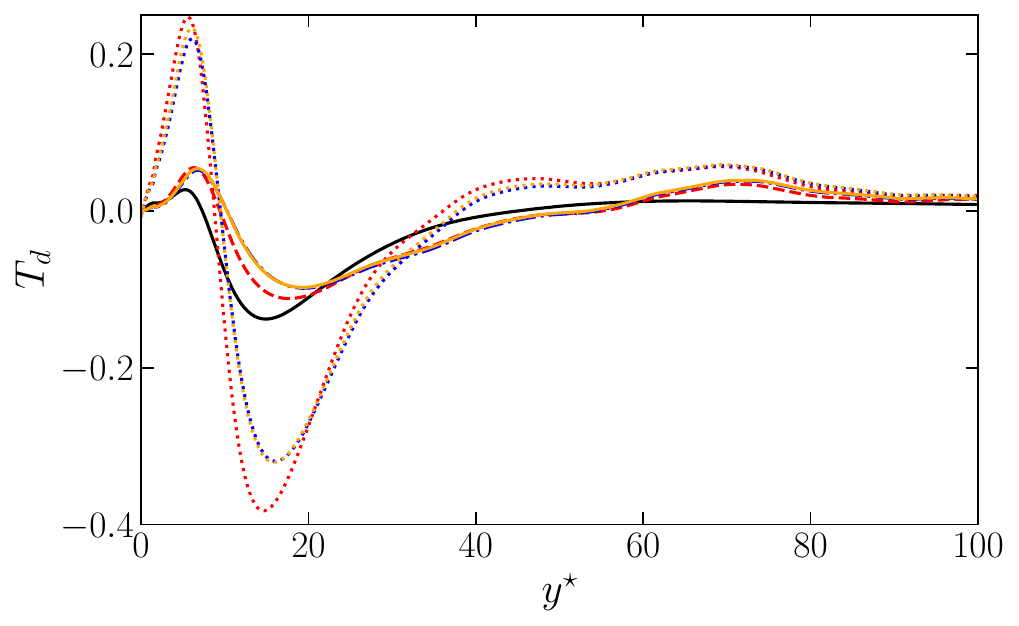}};
   \begin{scope}[x={(a.south east)},y={(a.north west)}]
     \node [align=center] at (0.03,0.95) {(d)};
   \end{scope}
 \end{tikzpicture}\\[-0.5cm]
\caption{Wall-normal distributions of the Turbulent transport of $k$. Comparison of the exact term with the models, considering the ``exact'' $\mu_t$ and the $\mu_t$ from $k$-$\omega$ model. Cases M2 (a), M6 (b), M10C (c) and M12 (b).}
\label{fig:Turbulenttke_transport}
\end{figure}
The turbulent diffusion or turbulent transport $T_d$ is traditionally modelled by means of a gradient approximation (GA). For high-$M$ flows, density gradients influence the logarithmic region of the velocity profile and must be accounted for in order to recover the correct near-wall behaviour. Therefore, a compressibility correction for density variations was initially proposed by Catris \& Aupoix \cite{catris2000density} and later refined by Otero \emph{et al.} \cite{otero2018turbulence}. The models read
\begin{equation}
\frac{\partial}{\partial x_j} \Big(\frac{1}{2} \overline{\rho u_i''u_i''u_j''} \Big) =
\begin{cases}
\frac{\partial}{\partial x_j}\Big(\frac{\mu_t}{\sigma_k}\frac{\partial k}{\partial x_j}\Big) & \text{GA}\\[8pt]
\frac{\partial}{\partial x_j}\Big(\frac{1}{\overline{\rho}} \frac{\mu_{\tau}}{\sigma_k}\frac{\partial \overline{\rho} k}{\partial x_j}\Big) & \text{Catris \& Aupoix \cite{catris2000density}}\\[8pt]
\frac{1}{{\sqrt{\overline{\rho}}}}\frac{\partial}{\partial x_j}\Big(\frac{1}{{\sqrt{\overline{\rho}}}}   \frac{\mu_{\tau}}{\sigma_k}\frac{\partial \overline{\rho} k}{\partial x_j}\Big) & \text{Otero \emph{et al.} \cite{otero2018turbulence}}
\end{cases}
\end{equation}
with $\sigma_k$ from the SST $k$-$\omega$ model. In figure~\ref{fig:Turbulenttke_transport}, we compare the three models against the exact term, considering the ``exact'' $\mu_t$ and the $\mu_t$ from $k$-$\omega$. In all cases, the use of the SST $\mu_t$ leads to larger deviations from the DNS, whereas the discrepancies are reduced when considering the exact turbulent viscosity, since they are only due to the model chosen for the diffusion term. For M2, all models in question show small differences, since the variations in the flow properties are relatively small. The discrepancies tend to increase at higher Mach numbers, with the uncorrected model underpredicting the peak in the buffer layer. Of note, the Catris-Aupoix and Otero models are essentially superposed for all cases.
For the M12 case, all the models predict the position of the first peak in the buffer layer but overestimate its magnitude. The second peak position and magnitude are not predicted accurately by any of the models.
Overall, based on a priori tests, the corrections do not seem to improve the agreement with the DNS significantly.

\subsubsection{Turbulent Dissipation}
\begin{figure}[!tb]
\centering
\begin{tikzpicture}
   \node[anchor=south west,inner sep=0] (a) at (0,0) {\includegraphics[width=0.49\textwidth]{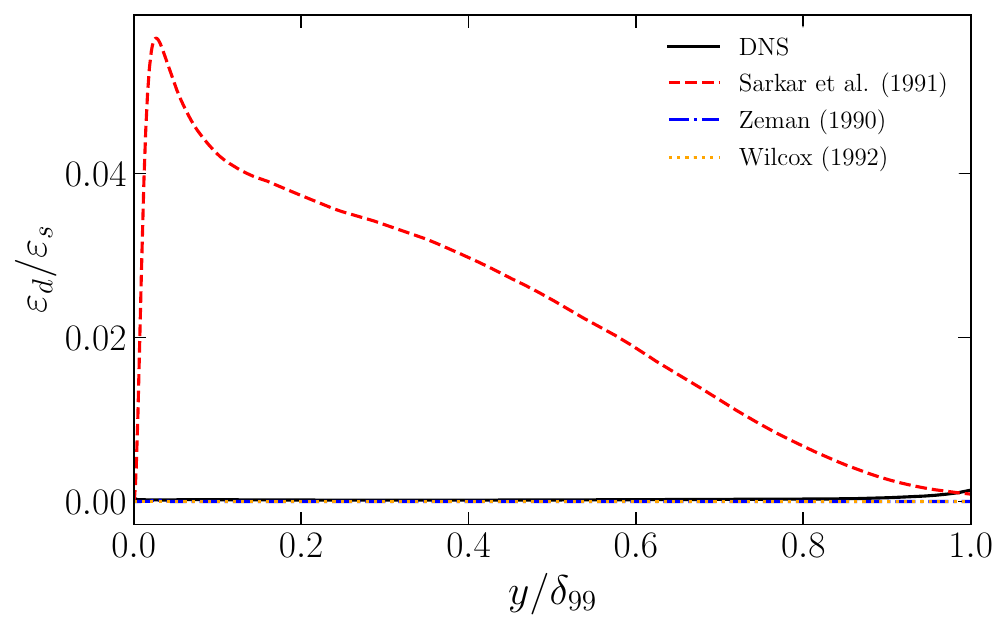}};
   \begin{scope}[x={(a.south east)},y={(a.north west)}]
     \node [align=center] at (0.03,0.95) {(a)};
   \end{scope}
 \end{tikzpicture}
 \hspace{-0.3cm}
 \begin{tikzpicture}
   \node[anchor=south west,inner sep=0] (a) at (0,0) {\includegraphics[width=0.49\textwidth]{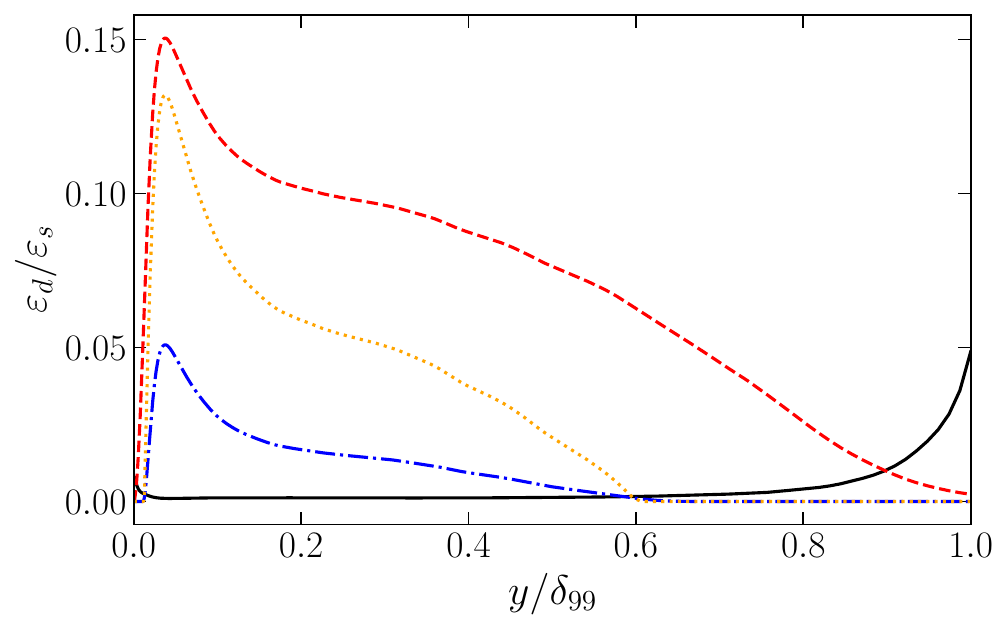}};
   \begin{scope}[x={(a.south east)},y={(a.north west)}]
     \node [align=center] at (0.03,0.95) {(b)};
   \end{scope}
 \end{tikzpicture}\\[-0.2cm]
 \begin{tikzpicture}
   \node[anchor=south west,inner sep=0] (a) at (0,0) {\includegraphics[width=0.49\textwidth]{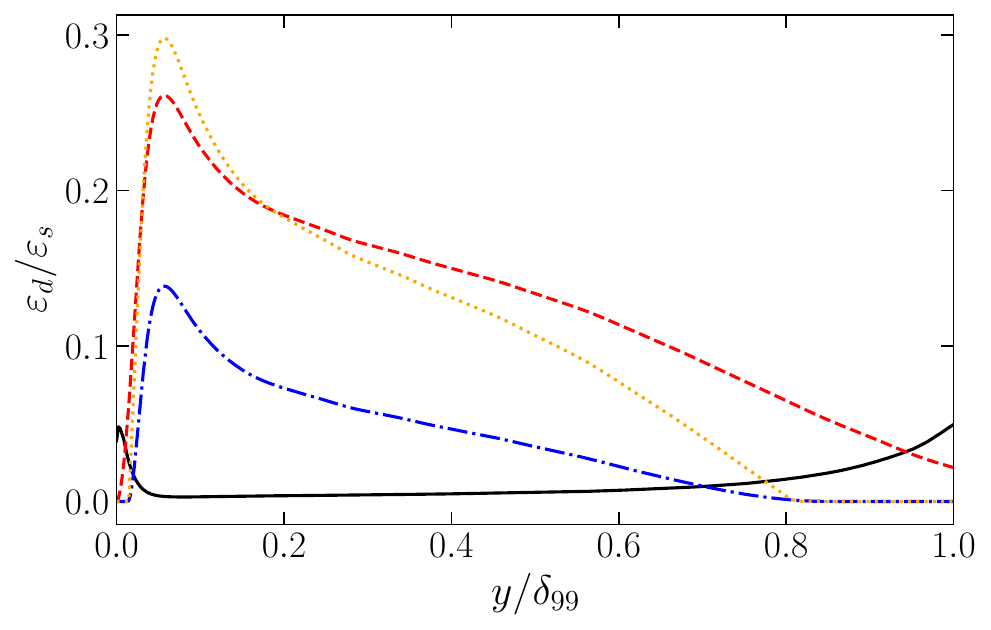}};
   \begin{scope}[x={(a.south east)},y={(a.north west)}]
     \node [align=center] at (0.03,0.95) {(c)};
   \end{scope}
 \end{tikzpicture}
 \hspace{-0.3cm}
 \begin{tikzpicture}
   \node[anchor=south west,inner sep=0] (a) at (0,0) {\includegraphics[width=0.49\textwidth]{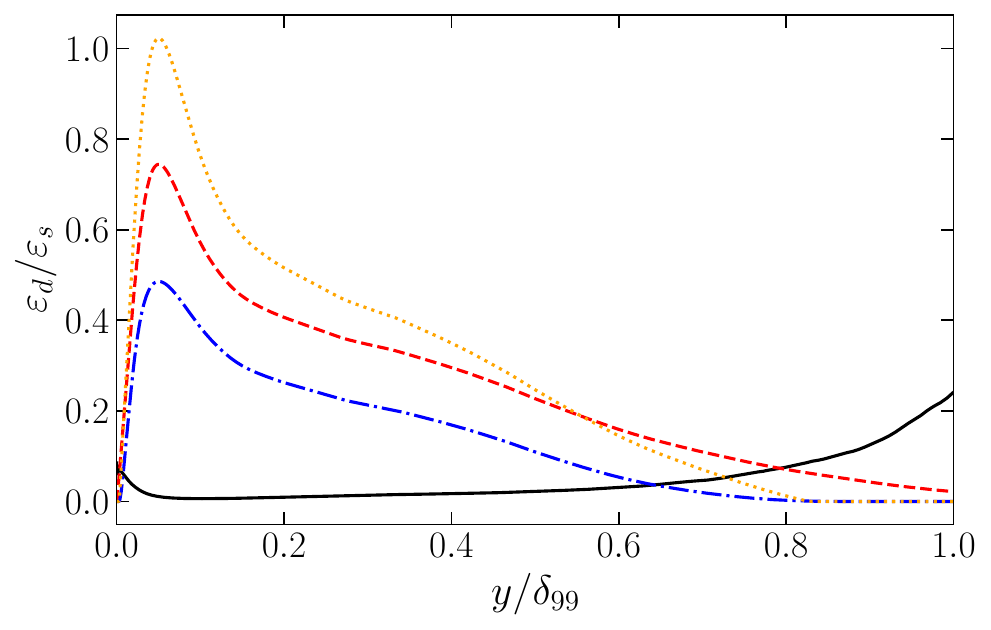}};
   \begin{scope}[x={(a.south east)},y={(a.north west)}]
     \node [align=center] at (0.03,0.95) {(d)};
   \end{scope}
 \end{tikzpicture}\\[-0.5cm]
\caption{Wall-normal distributions of the exact $\varepsilon_d/\varepsilon_s$ compared with the models listed in equation~\eqref{eq:dil_model}. Cases M2 (a), M6 (b), M10C (c) and M12 (d).}
\label{fig:epsilond_model}
\end{figure}
We examine next the turbulent kinetic energy dissipation $\varepsilon$. When dealing with compressible flows, it can be decomposed in three contributions:
\begin{equation}
    \overline{\rho} \varepsilon  = \overline{\rho} \varepsilon_s + \overline{\rho} \varepsilon_d + \overline{\rho} \varepsilon_i = \overline{\mu} \, \overline{ \omega_{i}' \omega_{i}'} + \frac{4}{3} \overline{\mu} \, \overline{ \vartheta'\vartheta'} + 2 \overline{\mu} \left[\frac{\partial^2 (\overline{u_i'u_j'}) }{\partial x_i x_j} - 2 \frac{\partial}{\partial x_j} \left( \overline{u_j' \frac{\partial u_i'}{\partial x_i'}} \right) \right]
\end{equation}
$\omega_i$ and $\vartheta$ being the vorticity components and the velocity divergence, respectively. The solenoidal dissipation $\varepsilon_s$ coincides with the dissipation in the incompressible limit, whereas the dilatational dissipation $\varepsilon_d$ appears only in compressible flows. Note that the inhomogeneous term $\varepsilon_i$ is shown to account for less than 2\% of the total dissipation and therefore it will not be considered hereafter.

For a two-equations model, the transport equations have to be adjusted to account for $\varepsilon_d$ and this is done by considering a closure model for the term.
A few models for $\varepsilon_d$ have been proposed in the literature, mostly based on DNS databases for homogeneous isotropic turbulence or free-shear flows \cite{wilcox1992dilatation, zeman1990dilatation,sarkar1991analysis}.
In all cases, the dilatation dissipation is assumed to take the general form $\varepsilon_d = F(M_t)\varepsilon_s$ \cite{gatski2013compressibility}, with
\begin{equation}\label{eq:dil_model}
    F(M_t) =
    \begin{cases}
    M_t^2 & \text{Sarkar \emph{et al.}~\cite{sarkar1991analysis}} \\
    \frac{3}{4} \left\{ 1 - \exp\left[ - \frac{\gamma+1}{2}\left(\frac{M_t - M_{t0}}{0.66}\right)^2 \right] \right\} \mathcal{H}(M_t - M_{t0}) & \text{Zeman \cite{zeman1990dilatation}}\\
    \frac{3}{2} \left[ M_t^2 - M_{t0}^2 \right]\mathcal{H}(M_t - M_{t0}) & \text{Wilcox \cite{wilcox1992dilatation}}
    \end{cases}
\end{equation}
$M_t$ being the turbulent Mach number, $M_{t0} = 0.25 \sqrt{2/(\gamma+1)}$ (with $\gamma = \overline{c_p}/\overline{c_v}$) and $M_{t0} = 0.25$ for Zeman and Wilcox models, respectively. Figure~\ref{fig:epsilond_model} reports profiles in semi-local scaling of the exact $\varepsilon_d/\varepsilon_s$ ratio from the DNS and its modelled counterpart according to the above-mentioned models.
The analysis of such ratio from DNS allows to better quantify the magnitude of compressibility effects. $\varepsilon_d/\varepsilon_s$ assumes maximum values of the order of $\approx 4$-$5 \%$ close to the wall (up to 10\% for M10 and M12 cases); it rapidly decreases and then becomes larger at the edge of the boundary layer, where $\varepsilon_s$ decays faster. Globally, it is rather small and neglecting it can therefore be considered a reasonable assumption. This does not seem to hold true for M12, as the wall cooling increases the significance of compressibility effects with $ \varepsilon_d \approx 7$-$8 \% \varepsilon_s$. The results for the three models considered show poor performance for all cases. This is a consequence of the simple functional relation~\eqref{eq:dil_model} assuming proportionality to $\varepsilon_s$ only through a function of $M_t$, leading the models to peak where $M_t$ peaks. Overall, the values tend to be largely overpredicted and the discrepancies increase with $M_\infty$.
As reported by \cite{wilcox1992dilatation}, the role of models for $\varepsilon_d$ is mostly to increase the overall dissipation of the RANS model, but all of them fail to correctly describe all of its physical dependencies on flow parameters.

\subsection{Turbulent heat flux}
\begin{figure}[!tb]
\centering
\includegraphics[width=0.49\textwidth]{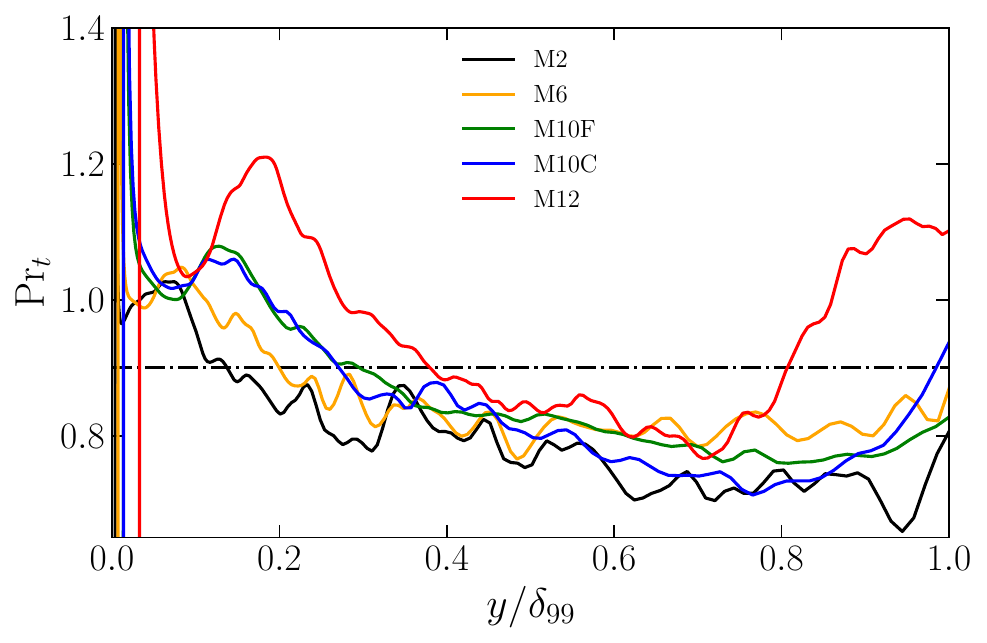}
\includegraphics[width=0.49\textwidth]{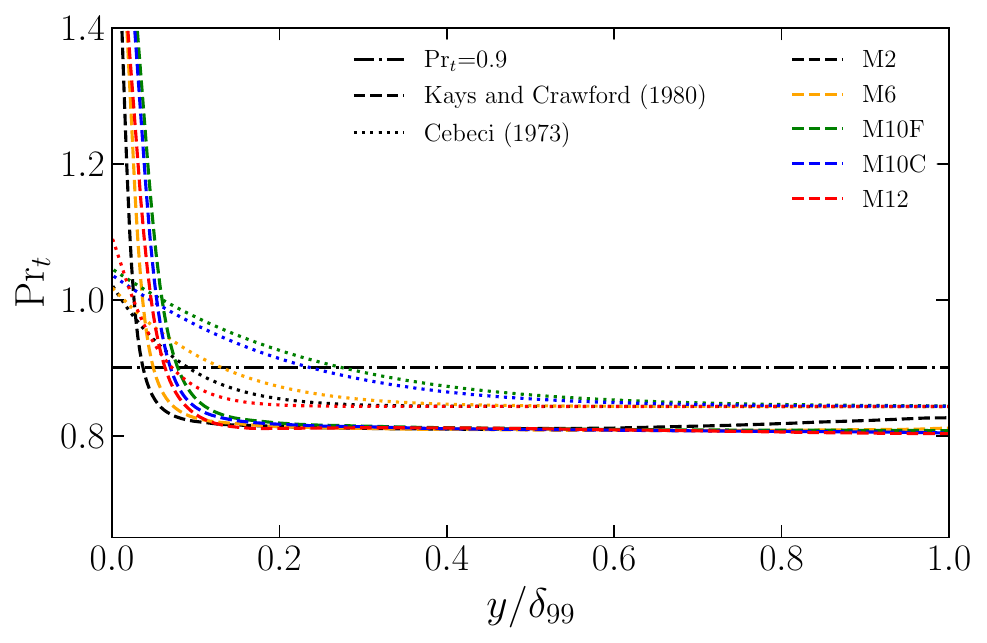}\\[-0.5cm]
\caption{Wall-normal distributions of the exact $\text{Pr}_t$ from DNS (left panel) and models from Kays-Crawford and Cebeci (right panel).
}
\label{fig:prt}
\end{figure}

In this section we assess the behavior of constitutive relations for the turbulent heat flux term $q^t_{j} = \overline{\rho u_j'' h''}$ that appears in the averaged total energy equation.
%
%
For the boundary layers at stake, we estimate an ``exact'' turbulent Prandtl number from the DNS data, as follows:
\begin{equation}
 \text{Pr}_t \approx \frac{\overline{\rho u''v'' }}{\rho v'' T''}\frac{\partial \widetilde{T}}{\partial y}\left(\frac{\partial \widetilde{u}}{\partial y}\right)^{-1}
\end{equation}
As mentioned, $\text{Pr}_t$ is most often assumed to be constant and equal to 0.9 for TBL.
More accurate results can be obtained by modelling $\text{Pr}_t$ and letting it vary; one example is the empirical relation of Kays and Crawford \cite{kays1980convective}:
\begin{equation}
    \text{Pr}_t = \left\{ \frac{1}{2 \text{Pr}_{t\infty}} +\frac{C \text{Pe}_t}{\sqrt{\text{Pr}_{t\infty}}} \right. \left. -(C \text{Pe}_t)^2 \left[1-\exp\left(-\frac{1}{C \text{Pe}_t \sqrt{\text{Pr}_{t\infty}} }\right) \right]   \right\}^{-1}
\end{equation}
where $C=0.3$, $\text{Pr}_{t\infty}$ is the free-stream value and $\text{Pe}_t = \text{Pr} \frac{\mu_t}{\overline{\mu}}$ is a turbulent Peclet number. Instead of using the classical value $\text{Pr}_{t\infty}=0.9$, we set it to 0.8. Another example is the model presented by Cebeci \cite{cebeci1973model}
\begin{equation}
\text{Pr}_{t} =\frac{\kappa}{\kappa_{\theta}} \frac{1- \exp\left(-y^{+} / A\right)}{1- \exp\left(-y^{+} / B\right)} \qquad \text{ with } \quad B =\frac{1}{\sqrt{\text{Pr}}} \sum_{i=1}^{5} C_{i}\left(\log_{10} \text{Pr}\right)^{i-1}
\end{equation}
with $\kappa = 0.4$, $\kappa_{\theta} = 0.44$, $A=26$, $C_{1}=34.96$, $C_{2}=28.79$, $C_{3}=33.95$, $C_{4}=6.3$, and $C_{5}=-1.186$. In Figure~\ref{fig:prt} the comparison between the exact $\text{Pr}_t$, the models and $\text{Pr}_t = 0.9$ is shown.
It appears that both the variable models and the constant assumption fail to accurately reproduce the trend of $\text{Pr}_t$ for all the cases considered. The Kays-Crawford model provides a better representation of the near wall behavior, but neither model captures the ``bump'' located at $y/\delta_{99}<0.2$, which is particularly accentuated for the M12 case. In the outer region, the exact values are below 0.9.

\subsection{Mass transport and turbulence-chemistry interaction}
\label{sec:mass}
In this section we assess models of unclosed terms in the species transport equations. We first focus on the turbulent mass transport term $\overline{\rho u_{j}^{\prime \prime} Y_{n}^{\prime \prime}}$ (n being the number of species) for a turbulent boundary layer, and we plot the ``exact'' Schmidt numbers,
computed from DNS data as:
\begin{align}
\quad Sc_{t,n}&=\frac{\overline{\rho u^{\prime \prime} v^{\prime \prime}} \partial \widetilde{Y}_{n} / \partial y}{\overline{\rho v^{\prime \prime} Y_{n}^{\prime \prime}} \partial \widetilde{u} / \partial y}
\end{align}
\begin{figure}[!tb]
\centering
\includegraphics[width=0.5\textwidth]{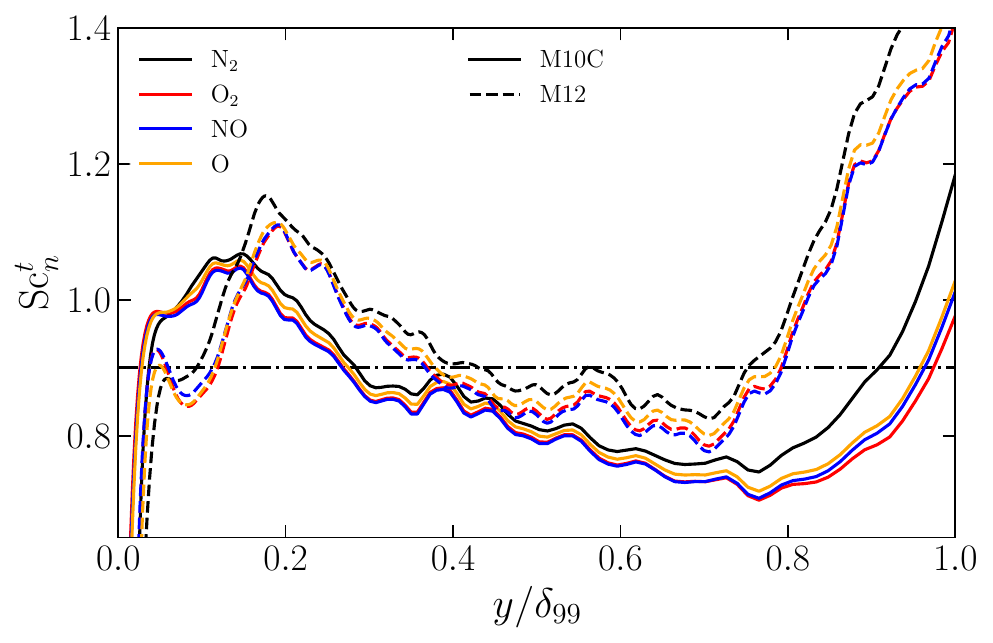}\\[-0.5cm]
\caption{Wall-normal distributions of $Sc_t$ from DNS for cases M10C and M12.
}
\label{fig:sct}
\end{figure}
The results are reported in figure \ref{fig:sct} for the species considered in the M10C and M12 cases.
Similar to $Pr_t$, the $Sc_{t,n}$ exhibit a bump in the logarithmic region, more pronounced for M12, and are below the standard value in the outer region. The value of 0.9  can then be considered as an average across the boundary layer. The profiles for different species do not differ much but their values and the position of the bump in the logarithmic region are clearly dependent on $M_\infty$.

\begin{figure}[!tb]
\centering
\includegraphics[width=0.49\textwidth]{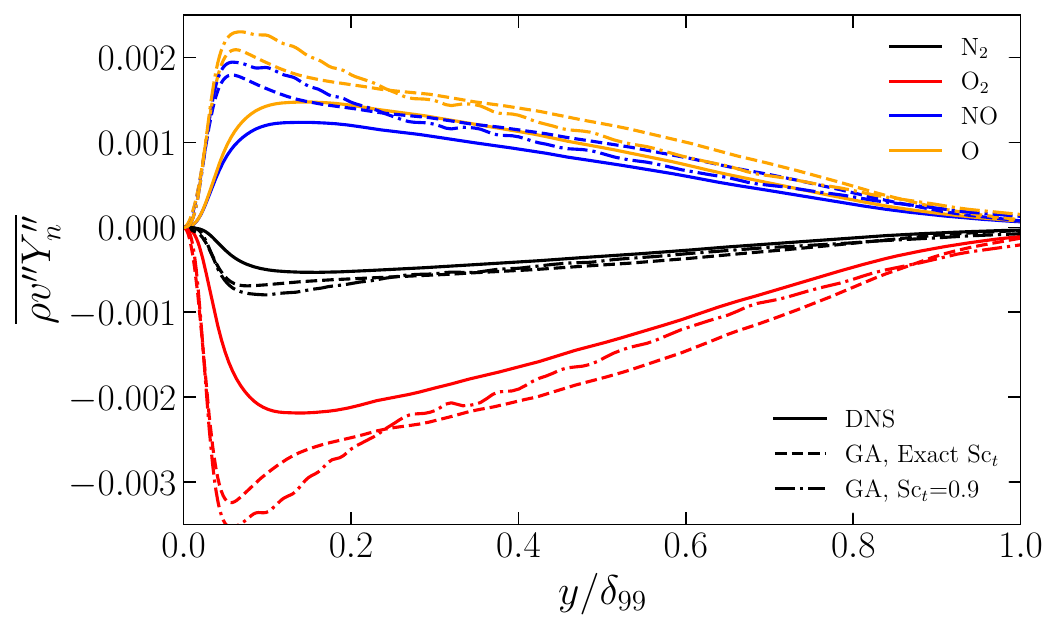}
\includegraphics[width=0.49\textwidth]{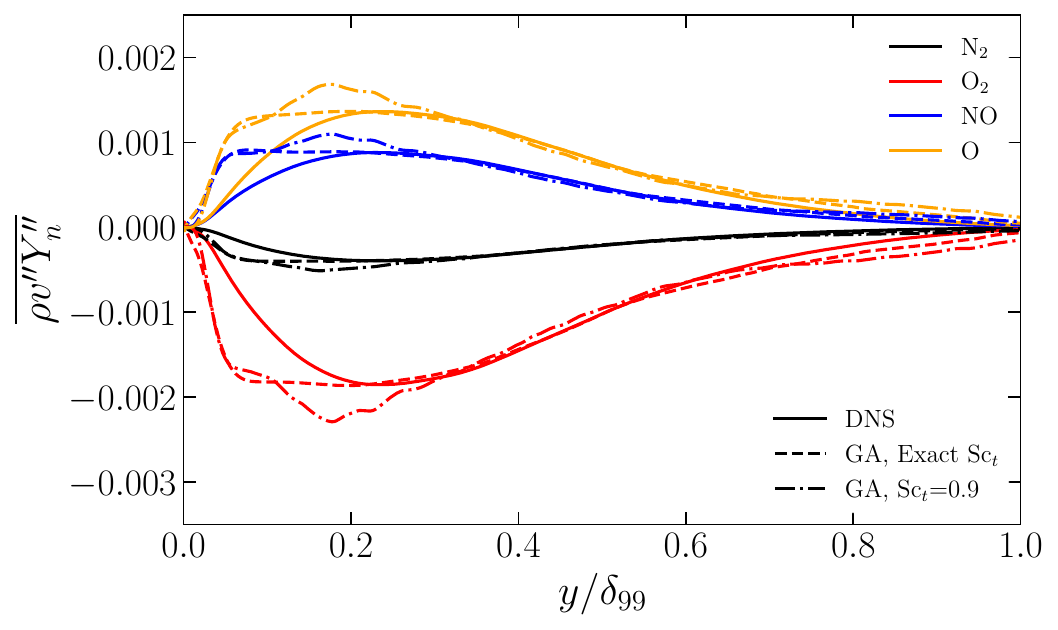}\\[-0.5cm]
\caption{Wall-normal profiles of the species turbulent fluxes for the M10C (left panel) and M12 (right panel) simulations.}
\label{fig:turbulent_species_flux}
\end{figure}
Figure~\ref{fig:turbulent_species_flux} shows the profiles of the turbulent species flux for the M10C and M12 simulations, using the classical gradient diffusion model $\overline{\rho v'' Y_n''} = - \frac{\mu_t}{\text{Sc}^t_n} \frac{\partial\widetilde{Y}_n}{\partial y}$ (named GA in the figures) with both the exact Sc$_t$ and Sc$_t=0.9$. Deviations of the turbulent Schmidt numbers from the constant 0.9 affect only moderately the resulting modeled fluxes, which are in reasonable agreement with the corresponding DNS statistics for $y/\delta_{99}\gtrsim 0.4$. Larger deviations are observed in the near-wall region, which are more related to the GA than to the constant Schimidt number hypothesis. This observation calls for the development of more sophisticated models of turbulent species transport, especially for problems where near wall mass exchange is important (e.g., for catalysis or ablation).\\

\begin{figure}[!tb]
\centering
\includegraphics[width=0.49\textwidth]{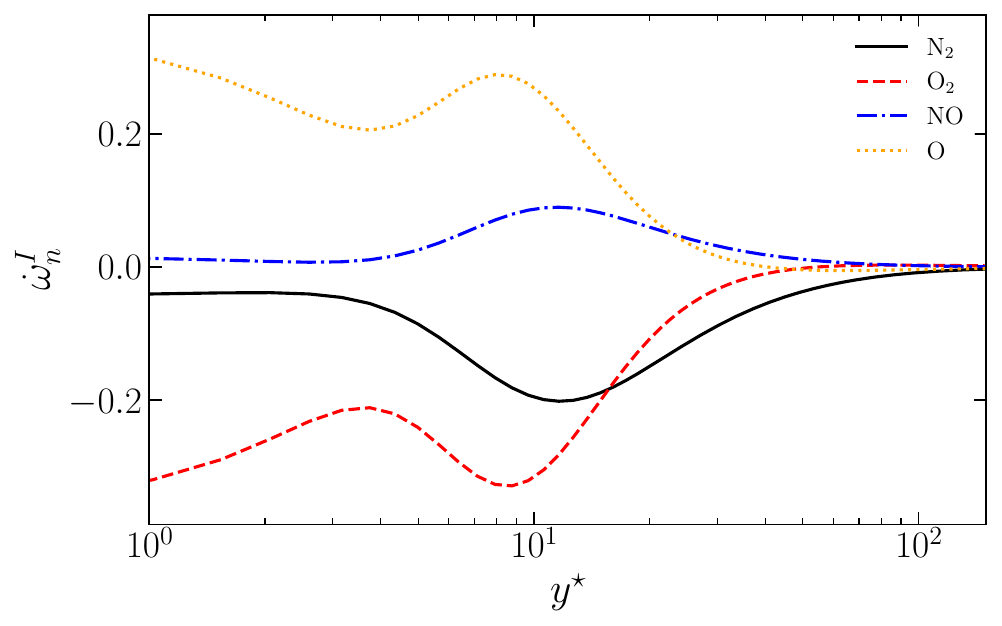}
\includegraphics[width=0.49\textwidth]{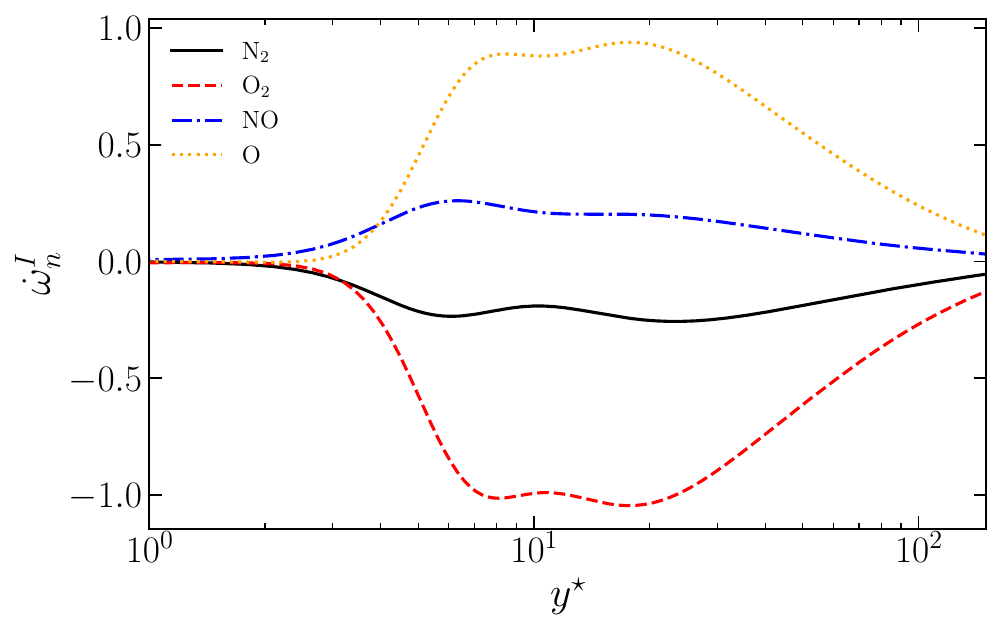}\\[-0.5cm]
\caption{Wall-normal profiles of the species source term fluctuations for the M10C (left panel) and M12 (right panel) simulations.}
\label{fig:dahmkoler_species}
\end{figure}
\begin{figure}[!tb]
\centering
\includegraphics[width=0.49\textwidth]{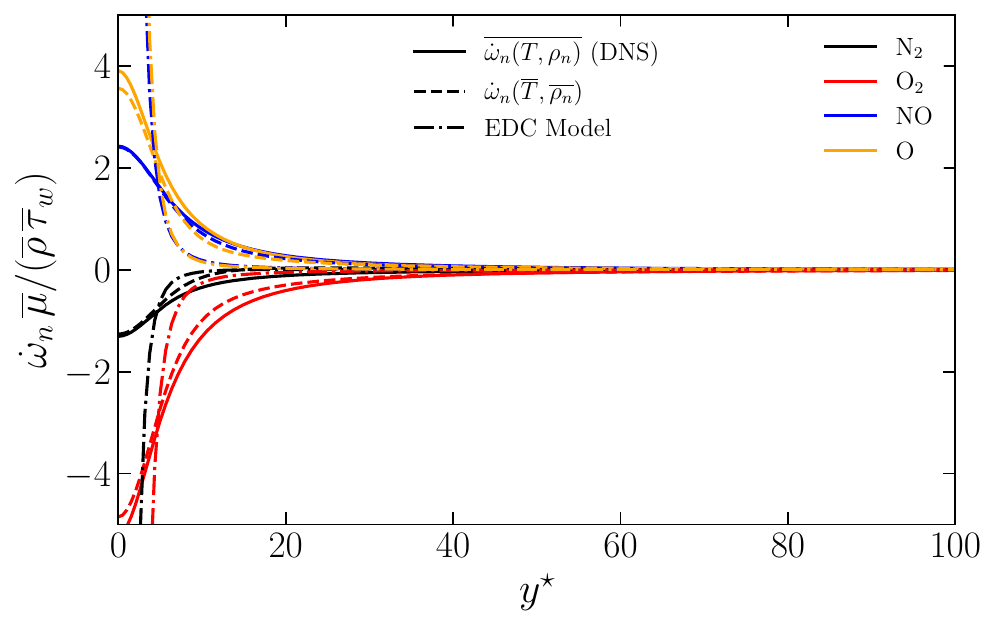}
\includegraphics[width=0.49\textwidth]{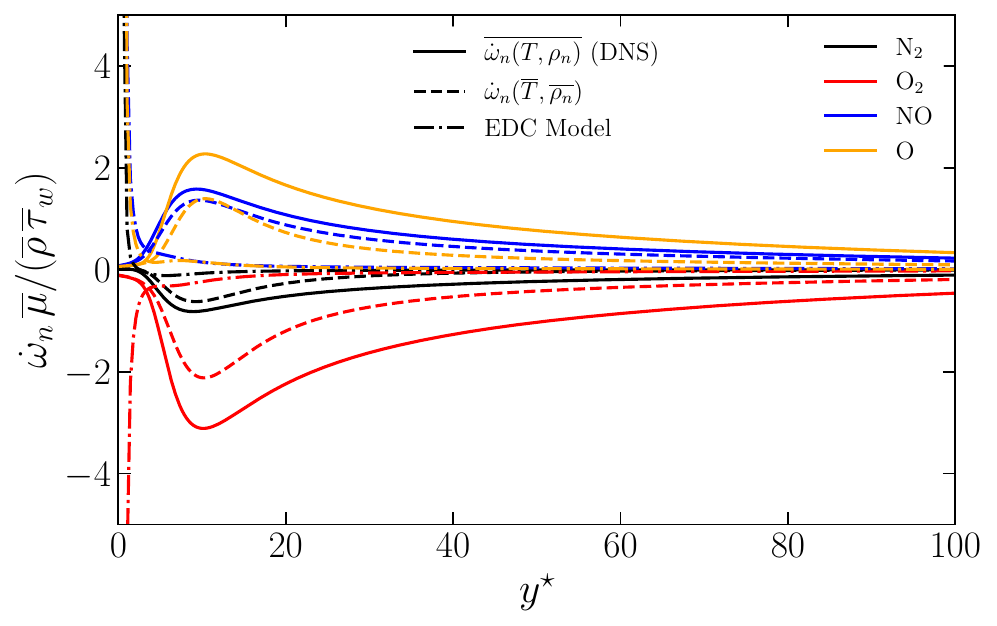}\\[-0.5cm]
\caption{Wall-normal profiles of the normalized species source term models for M10C (left panel) and M12 (right panel).}
\label{fig:omega_models}
\end{figure}
The influence of species mass fraction and temperature fluctuations on the species production rates, i.e. the intensity of turbulence-chemistry interactions (TCI), is evaluated through the quantity $\dot{\omega}_n^{I} = \left[\overline{\dot{\omega}_{n}\left(T, \rho_{n}\right)}-\dot{\omega}_{n}(\overline{T}, \overline{\rho_{n}})\right] \frac{\overline{\mu}}{\overline{\rho} \, \overline{\tau}_w}$
%
%
where $\overline{\dot{\omega}_n(T,\rho_n)}$ is a source term which represents the production or depletion of the $n$-th species in the mixture due to chemical reactions.
This parameter represents the chemical production due to turbulent fluctuations caused by the non zero difference between $\overline{\dot{\omega}_n(T,\rho_n)}$ and $\dot{\omega}_n(\overline{T},\overline{\rho_n})$, since the expression for $\dot{\omega}_n$ is highly nonlinear.
A simplification often adopted in the RANS framework is to consider the laminar closure $\overline{\dot{\omega}_n(T,\rho_n}) \approx \dot{\omega}_n(\overline{T},\overline{\rho_n})$, which is an acceptable approximation only if the turbulence-chemistry interactions are limited.
Due to the lack of models for the source terms in hypersonic TBL, an approach used for combustion applications is tested. We consider the model of Xiang \emph{et al.} \cite{xiang2020turbulence}, based on the Eddy Dissipation Concept (EDC); the chemical source term is then expressed as:
\begin{equation}
    \overline{\dot{\omega}_n} \approx \gamma^{*} \dot{\omega}_n(\overline{T},\overline{\rho_n}).
\end{equation}
Here, $\gamma^{*}  \approx 9.7 (\nu \varepsilon / k^2)^{\frac{3}{4}}$ represents the fine-scale structure volume fraction, i.e. the fraction of the volume in which chemical reactions take place which is assumed to be in the region where turbulent kinetic energy is quasi-steady.
Figure~\ref{fig:dahmkoler_species} shows the profile of $\dot{\omega}_n^{I}$ for the different species of the mixture; N is not shown being present in negligible amounts. The indicator assumes large values close to the wall for the adiabatic case M10C, and in the buffer region for the wall-cooled case M12, where turbulent fluctuations are significant.
To gain a better overview of TCI, the profiles of $\overline{\dot{\omega}_{n}\left(T, \rho_{n}\right)}$, $\dot{\omega}_{n}\left(\overline{T}, \overline{\rho_{n}}\right)$ and the EDC model are shown in figure~\ref{fig:omega_models}.
It can be observed that the assumption $\overline{\dot{\omega}_n(T,\rho_n}) \approx \dot{\omega}_n(\overline{T},\overline{\rho_n})$ is reasonably accurate for the adiabatic case, but underestimates the DNS value in the cooled case. On the contrary, the EDC model does not prove to be satisfactory, mainly due to the isotropic assumption in the formulation of $\gamma^{*}$ causing the model to fail close to the wall. Further research is required on closures for this term, by leveraging for instance presumed PDF models typically used in turbulent combustion applications \cite{bray2006finite}.
%
%
\subsection{Vibrational energy transport and turbulence-relaxation interaction}
\label{sec:vibrational}
We complete the analysis by investigating the behavior of unclosed terms in the averaged vibrational energy equation.
\begin{figure}[!tb]
\centering
\includegraphics[width=0.49\textwidth]{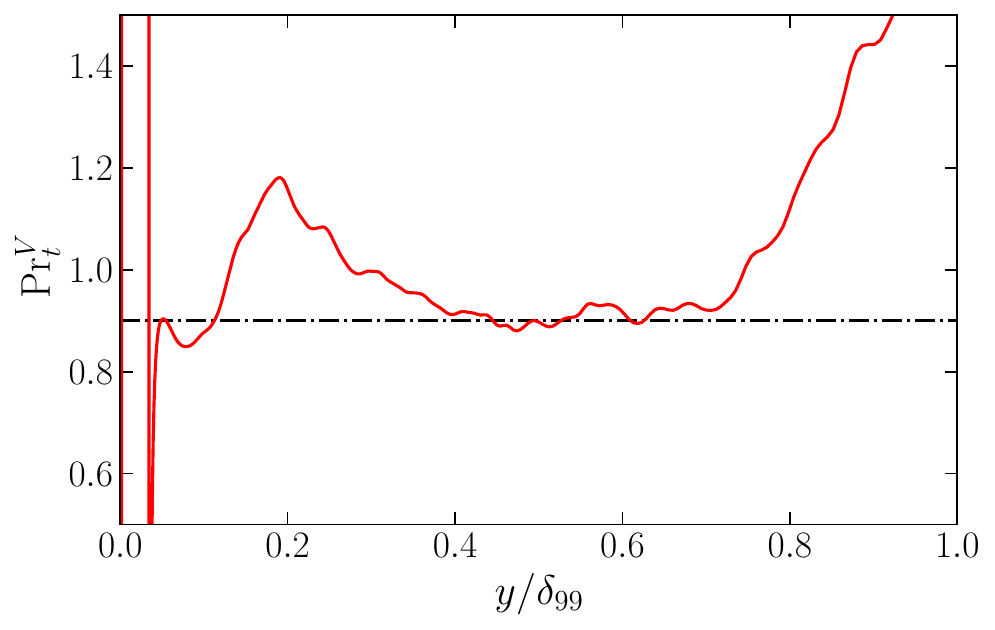}
\includegraphics[width=0.49\textwidth]{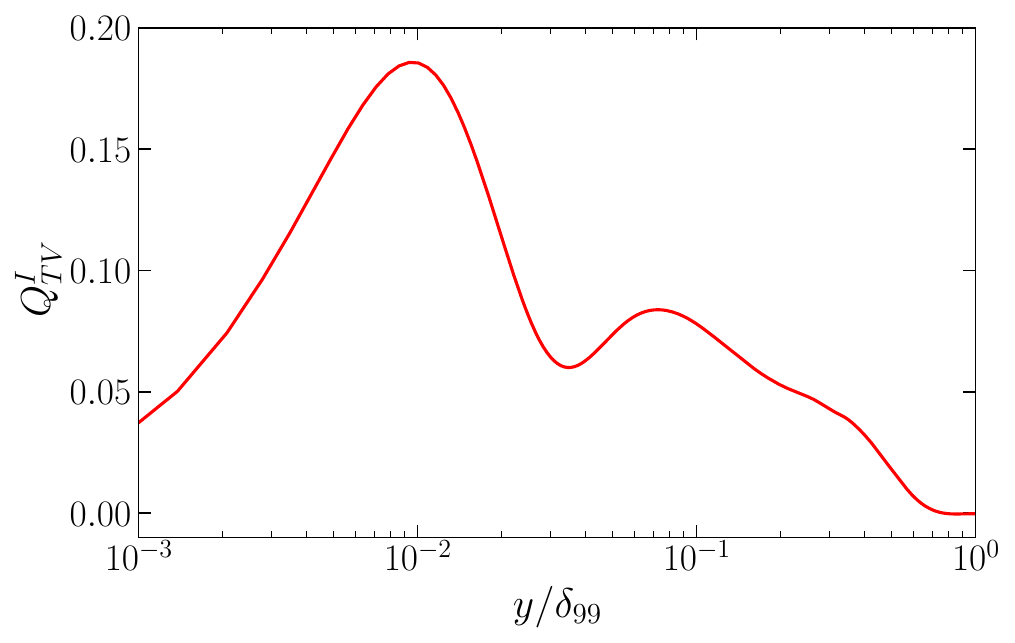}\\[-0.5cm]
\caption{(a) Exact $Pr_{t}^\text{V}$ from DNS and (b) source term interaction indicator for the M12 simulation.}
\label{fig:prvt_qtvi}
\end{figure}
As the heat flux was found to have an important part in the vibrational energy budget,  modeling this term appears critical in a RANS approach.
In \cite{passiatore2022thermochemical}, a gradient transport model is proposed, based on a vibrational turbulent Prandtl number $Pr_{t}^\text{V}$ defined in \cite{baranwal2020vibrational}.
Its ``exact'' value is extracted from DNS as:
\begin{equation}
    Pr_{t}^\text{V}=\frac{\overline{\rho u^{\prime \prime} v^{\prime \prime}} \partial \widetilde{T}_{\mathrm{V}} / \partial y}{\overline{\rho v^{\prime \prime} T_{\mathrm{V}}^{\prime \prime}} \partial \widetilde{u} / \partial y},
\end{equation}
such that the model reads:
\begin{equation}
    \overline{\rho u_{j}^{\prime \prime} e_{V}^{\prime \prime}} = \frac{\mu_t}{Pr_{t}^\text{V}}\frac{\partial \widetilde{e}_V}{\partial x_j}.
\end{equation}
The plot of figure~\ref{fig:prvt_qtvi}(a) shows a behavior similar to $Pr_t$ with a bump in the logarithmic region. In this case, the value 0.9 is reasonably well recovered in the outer region.
%
%
Finally we assess the intensity of the interaction between turbulence and thermal relaxation, through the indicator introduced in \cite{passiatore2022thermochemical}:
\begin{equation}
    Q_{\mathrm{TV}}^I=\frac{\overline{Q_{\mathrm{TV}}\left(T, T_{\mathrm{V}}, \rho, p, Y_n\right)}-Q_{\mathrm{TV}}\left(\widetilde{T}, \widetilde{T_{\mathrm{V}}}, \overline{\rho}, \overline{p}, \widetilde{Y_n}\right)}{\overline{Q_{\mathrm{TV}}^\text{max}}},
\end{equation}
where $\overline{Q_{\mathrm{TV}}}$ is a source term accounting for vibrational energy
production/depletion due to translational-vibrational energy transfers and $\overline{Q_{\mathrm{TV}}^\text{max}}$ is the maximum wall-normal value of at the selected station.
Figure~\ref{fig:prvt_qtvi}(b) shows differences up to $\approx 15\%$ in the inner layer, indicating a strong interaction between turbulence and thermal non-equilibrium.
While for $\dot{\omega}$ a few models do exist, allowing to quantify the intensity of TCI, no models for $\overline{Q_{\mathrm{TV}}}$ are available, which is potentially critical for RANS modeling of hypersonic flows out of thermal equilibrium.
\section{Conclusions}\label{sec:conclusions}
We reported a priori tests of RANS closures and compressibility corrections available in literature for various TBL configurations. Reference data were extracted from DNS databases ranging from the supersonic to the hypersonic regime,
including conditions of thermochemical non equilibrium.
Direct extensions of incompressible closures for unclosed terms in the turbulent kinetic energy equation are show to be inadequate at the considered Mach numbers. Unfortunately, the few compressibility corrections available in the literature are not found to capture DNS data much better, especially at the highest Mach numbers and in presence of wall cooling. While existing compressibility corrections for the turbulent diffusion of turbulent kinetic energy slightly improve the agreement with DNS data, models for the dilatational dissipation of turbulent kinetic energy introduce large errors, due to the hypothesis of a direct link with the local turbulent Mach number used in all such models. Fortunately, the dilatational dissipation rarely exceeds 4 to 5\% of the total dissipation; in addition, its contribution is limited to the inner region. As a consequence, neglecting its contribution seems a better choice than using any of the models available.

The gradient approximation commonly used to model turbulent heat fluxes, was found unable to predict the stream-wise component of various terms, pointing out limitations in the Reynolds analogy, on which the model is based.
Using non-constant turbulent Prandtl number models was not shown to improve accuracy.
Other closure approaches specific to this term should be developed.
High-temperature effects such as chemical or thermochemical nonequilibrium were found not to alter the validity of classical closure models. However, new unclosed terms appear in the species and vibrational energy equations need to be modeled accordingly.
Turbulent mass transport terms in the transport equations for chemical species were shown not to follow a gradient approximations, and the discrepancies cannot be amended by simply using a non-constant turbulent Schmidt number. Similar behaviors were observed independently of the chemical species at stake, with large discrepancies in the near-wall region, while the turbulent mass transport flux in the outer region is well approximated by the gradient model with $Sc_{t,n}=0,9$.
Models for the vibrational turbulent heat flux, adapted from the ones for turbulent heat fluxes, were shown to provide acceptable results.

 Overall, the study reveals that further work on many aspects of compressible turbulence modeling is required, in particular in the development of closures that account for genuine compressibility effects and shock/boundary layer interactions. A promising approach seems to be given by model machine learning techniques, e.g., such as those successfully employed in \cite{schmelzer2020discovery,cherroud2022sparse} for incompressible separated flows. This will make the object of future research.


\printbibliography

\end{document}